\documentclass{emulateapj}
\usepackage{graphicx}

\newcommand{\mum}{\ifmmode{\rm \mu m}\else{$\mu$m}\fi}

\begin{document}

\title{Carbon-rich dust production in metal-poor galaxies in the Local Group}


\author{
G.~C.~Sloan\altaffilmark{1}, 
M.~Matsuura\altaffilmark{2},
E.~Lagadec\altaffilmark{3},
J.~Th.~van~Loon\altaffilmark{4},
K.~E.~Kraemer\altaffilmark{5},
I.~McDonald\altaffilmark{6},
M.~A.~T.~Groenewegen\altaffilmark{7}
P.~R.~Wood\altaffilmark{8},
J.~Bernard-Salas\altaffilmark{9},
\& A.~A.~Zijlstra\altaffilmark{6}
}
\altaffiltext{1}{Cornell University, Astronomy Department,
  Ithaca, NY 14853-6801, USA, sloan@isc.astro.cornell.edu}
\altaffiltext{2}{Astrophysics Group, Department of Physics and Astronomy, 
  University College London, Gower Street, London WC1E 6BT, United Kingdom}
\altaffiltext{3}{European Southern Observatory, Karl 
  Schwarzschildstrasse 2, Garching 85748, Germany}
\altaffiltext{4}{Lennard-Jones Laboratories, Keele University, 
  Staffordshire ST5 5BG, United Kingdom}
\altaffiltext{5}{Institute for Scientific Research, Boston College,
  140 Commonwealth Avenue, Chestnut Hill, MA 02467, USA}
\altaffiltext{6}{Jodrell Bank Centre for Astrophysics, School 
  of Physics \& Astronomy, University of Manchester, Oxford 
  Road, Manchester M13 9PL, United Kingdom}
\altaffiltext{7}{Koninklijke Sterrenwacht van Belgi\"{e}, Ringlaan 3, 
  1180 Brussels, Belgium}
\altaffiltext{8}{Research School of Astronomy \& Astrophysics,
  Australian National University, Weston Creek, ACT 2611, Australia}
\altaffiltext{9}{Institut d'Astrophysique Spatiale, CNRS/Universite 
  Paris-Sud 11, 91405 Orsay, France}

\slugcomment{Submitted to Astrophysical Journal, 2011 Dec 22, Accepted
2012 Apr 24}


\begin{abstract}
We have observed a sample of 19 carbon stars in the Sculptor,
Carina, Fornax, and Leo~I dwarf spheroidal galaxies with the
Infrared Spectrograph on the {\it Spitzer Space Telescope.}  
The spectra show significant quantities of dust around the 
carbon stars in Sculptor, Fornax, and Leo~I, but little in
Carina.  Previous comparisons of carbon stars with similar 
pulsation properties in the Galaxy and the Magellanic Clouds 
revealed no evidence that metallicity affected the production
of dust by carbon stars.  However, the more metal-poor stars 
in the current sample appear to be generating less dust.  
These data extend two known trends to lower metallicities.  
In more metal-poor samples, the SiC dust emission weakens, 
while the acetylene absorption strengthens.  The bolometric 
magnitudes and infrared spectral properties of the carbon 
stars in Fornax are consistent with metallicities more 
similar to carbon stars in the Magellanic Clouds than in the 
other dwarf spheroidals in our sample.  A study of the carbon 
budget in these stars reinforces previous considerations that 
the dredge-up of sufficient quantities of carbon from the 
stellar cores may trigger the final superwind phase, ending a 
star's lifetime on the asymptotic giant branch.
\end{abstract}


\section{Introduction} 

Stars on the asymptotic giant branch (AGB) are an important
source of dust injected into the interstellar medium in the
Milky Way \citep[e.g.,][]{geh89, hab96}.  How important they
are in more metal-poor environments is an open question, with 
consequences for the early history of the Milky Way, current 
conditions in other smaller Local Group galaxies, and even 
for galaxies in the high-redshift Universe.  

The sensitivity of the Infrared Spectrograph 
\citep[IRS;][]{hou04} on the {\it Spitzer Space Telescope} 
\citep{wer04} has made it possible to explore this question 
by observing the dust forming around individual evolved stars 
in environments spanning a range of metallicities, both in 
our own galaxy and elsewhere in the Local Group.  The IRS has 
observed dust around AGB stars in the Large Magellanic Cloud 
\citep[LMC;][]{zij06,buc06,lei08}, Small Magellanic Cloud 
\citep[SMC;][]{slo06,lag07}, Fornax Dwarf Spheroidal 
\citep{mat07}, Sagittarius Dwarf Spheroidal 
\citep{lag09}\footnote{Also known as the Sagittarius Dwarf 
Elliptical Galaxy, or SAGDEG, and not to be confused with the 
Sagittarius Dwarf Irregular Galaxy, or SAGDIG.}, Sculptor 
Dwarf Spheroidal \citep{slo09}, and several Galactic globular 
clusters \citep{leb06,slo10,mcd11}.  These metalliticies of
these systems span a range of $-$2.1 $<$ [Fe/H] $<$ 0.0.
Comparisons of these samples with each other and with samples 
from the Galactic disk reveal that the amount of dust 
produced around oxygen-rich AGB stars decreases in more 
metal-poor environments \citep{slo08,slo10}, but for 
carbon-rich AGB stars, the amount of dust observed shows no 
measurable dependence on metallicity 
\citep{mat07,gro07,slo08}.

Stars on the AGB will produce either carbon-dominated or
oxygen-dominated dust, depending on their dredge-up
histories and initial abundances.  AGB stars generate carbon 
via the triple-alpha sequence \citep{sal52} in a 
helium-burning shell around an inert C/O core 
\citep[e.g.,][]{ir83}.  The helium fusion proceeds in a 
series of thermal pulses in the interior, leading to pulses 
of convection which dredge newly produced carbon up to the 
surface.  The dredge-ups can raise the photospheric C/O ratio 
above unity, converting an AGB star with a spectral class of 
M giant to a carbon star.  The envelopes of AGB stars are 
unstable to pulsations with typical periods of hundreds of 
days, making them readily identifiable as long-period
variables (LPVs).  These pulsations in the envelopes may 
even drive the mass-loss process \citep[see][and references 
therein]{mat08}.  

The formation of CO in the resulting outflows will exhaust 
all of the available carbon or oxygen, whichever is less 
abundant, leading to a chemical dichotomy in the dust which 
will condense out of the outflowing gas.  Alumina and 
silicates will dominate the shells around M giants, and 
amorphous carbon will dominate the shells around carbon 
stars \citep[e.g.,][]{mr87,ona89,es01}.

In more metal-poor galaxies, stars of lower initial mass will 
become carbon stars on the AGB.  Counts of carbon stars in 
the LMC and SMC reveal this fact observationally 
\citep{bbm78,bbm80,ch03}, and it is expected theoretically 
\citep{rv81,kl07}.  A recent infrared census of the SMC with 
{\it Spitzer} reveals the consequence:  carbon stars produce
more dust than their oxygen-rich AGB counterparts or red
supergiants \citep{mat12}, possibly much more 
\citep{boy12}.\footnote{The contribution from SNe at this
time is highly uncertain due to contradictory measurements
at different wavelengths; see \citep{mat11} for the
possibility that SNe can produce large amounts of dust.}
The SMC serves as a proxy for metal-poor galaxies too distant 
for their constituent stars to be studied individually.  By 
studying even more metal-poor galaxies in the Local Group, we 
can push to even more primitive systems.

The infrared spectra of seven carbon stars beyond the
Magellanic Clouds have been published so far, six in Fornax
\citep{mat07}, and one in Sculptor \citep{slo09}.  This paper 
presents spectra from the IRS for a larger sample of carbon 
stars in these two galaxies, as well as the Carina and Leo~I 
dwarf spheroidal (dSph) galaxies.  The stars in the sample 
are quite faint in the infrared, and the development of a new 
algorithm to extract spectra from the two-dimensional IRS 
images \citep{lev10} makes it possible to analyze their
spectra similarly to closer and brighter samples.  
Additionally, near-infrared (NIR) monitoring from the 
South African Astronomical Observatory (SAAO) has provided 
information on the pulsation modes and periods of the 
targeted stars which was not available when {\it Spitzer} 
observed them.  These two improvements give us the
opportunity to extend the previous comparisons of mass loss 
and dust production in evolved stars to more distant 
galaxies with lower metallicities.

Section 2 presents our targeted galaxies, with an emphasis
on their distances and metallicities, and explains how we
selected our sample of stars.
Section 3 describes the observations and data reduction.
In Section 4, we determine bolometric magnitudes and use 
these to re-assess the metallicities.  Section 5 presents
the spectroscopic results, and Section 6 discusses the 
consequences of our findings.

\section{The sample} 

\subsection{Target galaxies\label{s.gal}} 

\begin{deluxetable*}{lrllc} 
\tablenum{1}
\tablecolumns{5}
\tablewidth{0pt}
\tablecaption{Dwarf Spheroidal Galaxies Studied}
\label{t.gal}
\tablehead{
  \colhead{dSph}     & \colhead{Distance} & 
  \colhead{Adopted}  & \colhead{Mean}     & \colhead{Adopted} \\
  \colhead{galaxy}   & \colhead{modulus}  & 
  \colhead{$E(B-V)$} & \colhead{[Fe/H]}   & \colhead{[Fe/H]}
}
\startdata
Sculptor & 19.64 $\pm$ 0.04 & 0.02  & $-$1.56 $\pm$ 0.40 & $\sim$$-$1.0 \tablenotemark{a} \\
Carina   & 20.10 $\pm$ 0.04 & 0.025 & $-$1.73 $\pm$ 0.35 & $-$1.73 \\
Fornax   & 20.74 $\pm$ 0.07 & 0.025 & $-$0.99 $\pm$ 0.44 & $-$0.3 to $-$0.8 \tablenotemark{a} \\
Leo I    & 22.07 $\pm$ 0.07 & 0.03  & $-$1.35 $\pm$ 0.24 & $-$1.35 
\enddata
\tablenotetext{a}{\S~\ref{s.mm} explains these revisions in
Sculptor and Fornax.}
\end{deluxetable*}

\begin{deluxetable}{lll} 
\tablenum{2}
\tablecolumns{3}
\tablewidth{0pt}
\tablecaption{Distances to the Galaxies}
\label{t.dm}
\tablehead{
  \colhead{Galaxy} & \colhead{Distance} & \colhead{Ref.}  \\
  \colhead{ }      & \colhead{modulus}  &  \colhead{ } 
}
\startdata
Sculptor & 19.71 $\pm$ 0.10 & \cite{kal95} \\
         & 19.64 $\pm$ 0.04 & \cite{riz07a} \\
         & 19.67 $\pm$ 0.12 & \cite{pie08} \\
\\
Carina   & 20.09 $\pm$ 0.06 & \cite{sme94} \\
         & 20.06 $\pm$ 0.12 & \cite{mat98} \\
         & 20.19 $\pm$ 0.12 & \cite{dal03} \\
         & 20.11 $\pm$ 0.13 & \cite{pie09} (avg.\ of $J$, $K$) \\
\\
Fornax   & 20.76 $\pm$ 0.10   & \cite{buo99} \\
         & 20.70 $\pm$ 0.12   & \cite{sav00} (tip of RGB) \\
         & 20.76 $\pm$ 0.04   & \cite{sav00} (HB) \\
         & 20.65 $\pm$ 0.11   & \cite{ber00} \\
         & 20.86 $\pm$ (0.04) & \cite{pie03} \\
         & 20.66 $\pm$ (0.04) & \cite{mg03} \\
         & 20.64 $\pm$ 0.09   & \cite{gre07} \\
         & 20.72 $\pm$ 0.04   & \cite{riz07b} \\
         & 20.75 $\pm$ 0.19   & \cite{gul07} (tip of RGB) \\
         & 20.75 $\pm$ 0.11   & \cite{gul07} (red clump) \\
         & 20.84 $\pm$ 0.14   & \cite{pie09} \\
\\ 
Leo I    & 22.18 $\pm$ 0.11 & \cite{lee93} \\
         & 22.00 $\pm$ 0.15 & \cite{cap99} \\
         & 22.04 $\pm$ 0.14 & \cite{hel01} \\
         & 22.05 $\pm$ 0.18 & \cite{mnd02} \\
         & 22.02 $\pm$ 0.13 & \cite{bel04} \\
         & 22.04 $\pm$ 0.11 & \cite{hel10} 
\enddata
\end{deluxetable}

Our targets sample the evolved stellar population in four
dwarf spheroidal galaxies in the Local Group.  Sculptor was 
the first dwarf spheroidal discovered, quickly followed by 
Fornax \citep{sha38}.  Leo~I was uncovered during the first 
Palomar Sky Survey \citep{hw50}, and \cite{can77} detected 
the Carina dwarf while conducting the Southern Sky Survey 
from the European Southern Observatory (ESO).  
Table~\ref{t.gal} presents some basic data for these galaxies 
that we will use for the remainder of the paper, and it 
requires some explanation.  

The distance moduli were determined with weighted averages of 
published distances based on standard candles such as 
RR~Lyrae variables, the horizontal branch (HB), and the tip 
of the red giant branch (RGB).  The uncertainties in the 
distance moduli are statistical and do not reflect systematic 
errors, which are likely to be larger.  Table~\ref{t.dm} 
lists the individual distance measurements used to determine 
the results in Table~\ref{t.gal}.  The entries in 
Table~\ref{t.dm} are not meant to be exhaustive; some 
measurements rendered redundant or obsolete by more recent 
work are not included.  Uncertainties smaller than 0.04 
magnitudes have been raised to that limit for the purpose of 
weighting (and noted with parentheses in Table~\ref{t.dm}).

Table~\ref{t.gal} includes our assumed values of interstellar 
reddening from foreground extinction in the Galaxy,
$E(B-V)$.  In this study, they only influence our derived 
bolometric magnitudes, and the influence is small, because the 
extinction is small to begin with and smaller still in the 
NIR, where the carbon stars emit most of their energy.  As an 
example, a reddening of 0.03 magnitudes corresponds to an 
extinction of 0.03 magnitudes at $J$ and 0.01 at $K$ 
\citep[using the extinction law of][]{rl85}, making the 
impact of reddening smaller than our uncertainty in distance.  
Our assumed reddening values are consistent with the 
literature and the infrared dust maps of \cite{sch98}, except 
that reddening derived from the dust maps is higher than 
typical values used for Carina \citep[0.06 vs.\ 0.025 mag.; 
e.g.,][]{sme94,mat98}.


\subsection{Metallicities\label{s.metal}} 

Our objective is to understand how the infrared spectral
characteristics of carbon stars vary with metallicity, making
it important that we understand the metallicity distribution
functions (MDFs) of the parent populations of our targeted 
carbon stars in each galaxy.  Tackling this question is not 
easy, given the complex star formation histories of these 
systems.  Each galaxy is unique in this regard.



\subsubsection{Sculptor}

Sculptor has two populations with different ages, 
metallicities, and spatial distributions.  Its 
color-magnitude diagram (CMD) shows two RGB bumps and two 
HBs, consistent with two populations with distinct 
metallicities \citep{maj99}.  \cite{hk99} noticed that the 
red HB, which arises from the more metal-rich population, is 
confined to the center of the galaxy.  

In the past decade, multiple studies have used the Fibre 
Large Multi-Element Spectrograph (FLAMES) at the Very Large 
Telescope (VLT), measuring the Ca II triplet (CaT) at 
0.85~\mum\ to determine the metallicity of hundreds of RGB 
stars.  Those that focus on the core regions (within 
$\sim$10--15$\arcmin$ of the center\footnote{All distances 
in this section are ellipsoidal, or foreshortened away from
the major axis.  See \cite{ih95} for geometrical details.}) 
tend to produce more metal-rich MDFs compared to those 
sampling larger parts of the galaxy.  Inside 12$\arcmin$ of 
the center, \cite{tol04} find a MDF with $<$[Fe/H]$>$ = 
$-$1.49 $\pm$ 0.35, but outside that ellipsoidal radius, 
$<$[Fe/H]$>$ = $-$1.91 $\pm$ 0.27 \citep[our calculations 
based on figures in their paper; all metallicities are on the 
CG97 scale from][]{cg97}.  Similarly, \cite{kir09} find 
$<$[Fe/H]$>$ = $-$1.58 $\pm$ 0.41 for a large sample within 
10$\arcmin$ of the center, while \cite{hel06} find 
$<$[Fe/H]$>$ = $-$1.82 $\pm$ 0.34 for a sample out to the 
tidal radius (76$\farcm$5; our $<$[Fe/H]$>$ calculation based 
on their data).  

\cite{rev09} explain this dichotomy with a young, metal-rich 
population (age $\la$ 2 Gyr), and an old metal-poor 
population (age $\ga$ 9 Gyr).  Their models indicate no 
intermediate-age population.  Our carbon stars are more 
likely to belong to the younger population (see 
\S~\ref{s.mm}), which corresponds to the population 
dominating the core of the galaxy.  To approximate the 
metallicity in the core, we have averaged the inner sample 
defined by \cite{tol04} and the sample of \cite{kir09}, 
weighting by sample size (97 and 393, respectively), to 
arrive at an estimated $<$[Fe/H]$>$ of $-$1.56 $\pm$ 0.40.
However, this value is too metal-poor compared to the 
metallicity of the younger population according to the 
models of \cite{rev09}, leading us to revise the 
metallicity in \S~\ref{s.mm}.


\subsubsection{Carina}

Carina has experienced multiple, discrete 
star-formation events \citep{mig90}.  \cite{sme96} detected 
three main-sequence turn-off points, along with three 
locations for helium-burning stars (two HBs and a red clump 
projected onto the RGB), which \cite{hk98} dated to three
star-formation events early in the galaxy's history and 
$\sim$7 and $\sim$3 Gyr ago.  This basic scenario has stood 
the test of another decade of observations \citep[e.g.,][and 
references therein]{mon03,bon10}.  Models by \cite{rev09} 
suggest that the intermediate population formed in a series 
of several bursts.  

Despite this complex star-formation history, Carina shows a 
relatively shallow metallicity gradient \citep{wal09} and a 
MDF no broader than that of Sculptor.  CaT observations with 
VLT/FLAMES by \cite{hel06} cover Carina out to 
$\sim$36$\arcmin$, and from their data we determine that 
$<$[Fe/H]$>$ = $-$1.81 $\pm$ 0.31.  Other recent estimates 
are $-$1.72 $\pm$ 0.39 \citep{koc06}, $-$1.69 $\pm$ 0.51 
\citep{koc08}, and $-$1.70 $\pm$ 0.19 \citep{bon10}.  The 
mean of these results is $-$1.73 $\pm$ 0.35 (we have
averaged the available uncertainties).


\subsubsection{Fornax}

Fornax has experienced a steadier rate of star 
formation over the past several Gyr compared to Sculptor and 
Carina, resulting in a broader MDF.  The CMD of Fornax has a 
wide RGB \citep{dem79}, which requires a range of 
metallicities and/or ages.  Fornax contains many carbon 
stars, proof of a substantial population of intermediate-age 
stars \citep{dk79,am80}.  Further study made it apparent that 
episodes of star formation have continued to within the last 
few hundred Myr \citep{am85,buo85,buo99}, and that younger 
stars are more concentrated in the core of the galaxy 
\citep{ste98}.  More recent work has largely confirmed these 
earlier findings \citep[e.g.,][]{tol01,pon04,hel06}.  

CaT observations in large samples of RGB stars by 
\cite{bat06} trace the metallicity gradient in Fornax.  They 
find a MDF within 24$\arcmin$ of the center with $<$[Fe/H]$>$ 
= $-$0.99 $\pm$ 0.44, compared to $-$1.52 $\pm$ 0.46 outside 
42$\arcmin$ of the center (the quoted quantities are our 
determinations from their data).  A study of CMDs within 
Fornax by \cite{cj08} support the spectroscopic results.  Our 
sample of carbon stars mostly conforms to the innermost 
sample considered by \cite{bat06}, and we will adopt $-$0.99 
as a starting metallicity for consideration.  

However, models by \cite{rev09} suggest that the carbon stars 
could be significantly more metal-rich; most of the stars 
formed in the past few Gyr should have [Fe/H] in the range 
from $\sim$ $-$0.3 to $-$0.8, which would make these carbon 
stars more similar to those in the LMC and SMC than in the 
other three dwarf spheroidal galaxies in our sample.  We 
return to this point in \S~\ref{s.mm} below.

\subsubsection{Leo~I}

Leo~I is the most distant of the dwarf galaxies in the 
Milky Way system of the Local Group.  In fact, it is 
unclear whether or not it is gravitationally bound to the 
Local Group \citep[e.g.,][]{lep11}.  Its distance and its 
proximity to Regulus have made observations more challenging 
than for the other dwarfs considered here.  Nonetheless, a 
picture has emerged of a galaxy with the contradictory 
properties of a relatively young population and relatively 
metal-poor abundances \citep{lee93}.  The majority of the 
visible stars in the galaxy appear to have formed 3--7 Gyr 
ago \citep{dem94} or 1--7 Gyr ago \citep{gal99}.  

Despite the ongoing star formation, gradients in the 
population are subtle \citep{gul09,hel10}, and the 
metallicity shows a fairly narrow and well-defined 
distribution.  CaT spectra give $<$[Fe/H]$>$ values of 
$-$1.34 $\pm$ 0.26 \citep[102 stars;][]{bos07}, $-$1.31 $\pm$ 
0.25 \citep[58 stars;][]{koc07}, and $-$1.41 $\pm$ 0.21 
\citep[54 stars;][]{gul09}.  Combining these results and 
weighting by their sample size yields $<$[Fe/H]$>$ = $-$1.35 
$\pm$ 0.24.

\subsection{Stellar samples} 

\begin{deluxetable*}{lllllll} 
\tablenum{3}
\tablecolumns{7}
\tablewidth{0pt}
\tablecaption{Targets}
\label{t.star}
\tablehead{
  \colhead{Source\tablenotemark{a}} & \multicolumn{2}{c}{Position (J2000)} & 
  \multicolumn{3}{c}{2MASS photometry} & \colhead{Other} \\
  \colhead{name} & \colhead{RA} & \colhead{Dec.} & \colhead{$J$} & 
  \colhead{$H$} & \colhead{$K_s$} & \colhead{designations\tablenotemark{a}}
}
\startdata
MAG 29        & 00 59 53.67 & $-$33 38 30.8 & 14.846 $\pm$ 0.038 & 13.144 $\pm$ 0.031 & 11.603 $\pm$ 0.021 \\
Scl V78 V544  & 00 59 58.94 & $-$33 28 35.2 & 13.399 $\pm$ 0.021 & 12.633 $\pm$ 0.025 & 12.273 $\pm$ 0.023 & ALW Scl 3\\
\\
For BW 2      & 02 38 06.19 & $-$34 31 19.4 & 16.052 $\pm$ 0.104 & 14.483 $\pm$ 0.055 & 13.315 $\pm$ 0.048 & GLM 31 \\
For BTH 13-23 & 02 38 50.56 & $-$34 40 32.0 & 16.106 $\pm$ 0.094 & 14.525 $\pm$ 0.053 & 12.879 $\pm$ 0.029 & \\
For BTH 12-4  & 02 39 12.33 & $-$34 32 45.0 & 14.722 $\pm$ 0.033 & 13.262 $\pm$ 0.038 & 12.120 $\pm$ 0.024 & GLM 25 \\
For BTH 3-129 & 02 39 41.60 & $-$34 35 56.7 & \nodata            & 15.970 $\pm$ 0.205 & 14.164 $\pm$ 0.070 & \\
For DK 18     & 02 39 54.21 & $-$34 38 36.9 & 15.601 $\pm$ 0.064 & 14.162 $\pm$ 0.049 & 13.167 $\pm$ 0.033 & GLM 24 \\
For DK 52     & 02 40 06.66 & $-$34 23 22.3 & 14.485 $\pm$ 0.029 & 13.377 $\pm$ 0.034 & 12.618 $\pm$ 0.027 & DDB 17, GLM 13 \\
For DI 2      & 02 40 09.47 & $-$34 06 25.7 & 15.790 $\pm$ 0.075 & 14.556 $\pm$ 0.068 & 13.668 $\pm$ 0.052 & GLM 16 \\
For WEL C10   & 02 40 10.17 & $-$34 33 21.9 & 14.063 $\pm$ 0.026 & 13.122 $\pm$ 0.021 & 12.545 $\pm$ 0.029 & DI 20, SHS 105, BW 62, DDB 19, BTH 4-25\\
For BW 69     & 02 40 17.79 & $-$34 27 35.8 & 15.424 $\pm$ 0.063 & 14.122 $\pm$ 0.049 & 13.182 $\pm$ 0.035 & GLM 21 \\
For BW 75     & 02 40 31.23 & $-$34 28 44.2 & 14.745 $\pm$ 0.043 & 13.689 $\pm$ 0.046 & 13.072 $\pm$ 0.037 & DDB 22, BTH 6-13, GLM 17 \\
For BW 83     & 02 41 03.56 & $-$34 48 05.4 & 14.441 $\pm$ 0.035 & 13.365 $\pm$ 0.034 & 12.694 $\pm$ 0.034 & DDB 25, MAG 30, GLM 27 \\
\\
ALW Car 2     & 06 41 13.53 & $-$50 54 25.0 & 13.925 $\pm$ 0.024 & 13.073 $\pm$ 0.028 & 12.658 $\pm$ 0.027
\\
Car MCA C3    & 06 41 41.45 & $-$50 58 08.1 & 13.742 $\pm$ 0.022 & 12.805 $\pm$ 0.023 & 12.340 $\pm$ 0.026 & ALW Car 6 \\
Car MCA C5    & 06 42 10.35 & $-$50 56 24.0 & 13.940 $\pm$ 0.023 & 13.159 $\pm$ 0.027 & 12.785 $\pm$ 0.029 & ALW Car 10 \\
\\
\\
Leo I MFT C   & 10 08 22.25 & $+$12 17 57.1 & \nodata            & 16.195 $\pm$ 0.214 & 14.225 $\pm$ 0.060 & HGR 8717 \\
Leo I MFT A   & 10 08 29.28 & $+$12 18 51.6 & 17.134 $\pm$ 0.202 & 15.429 $\pm$ 0.103 & 14.025 $\pm$ 0.053 & HGR 6343 \\
Leo I MFT E   & 10 09 00.5  & $+$12 19 01   & \nodata            & \nodata            & \nodata            
\enddata
\tablenotetext{a}{V78 = \cite{va78}; DK = \cite{dk79}; MCA = \cite{mou82};
ALW = \cite{alw85,alw86}; WEL = \cite{wes87}; DI = \cite{di87}; 
SHS = \cite{ste98}; BW = \cite{bw02}; DDB = \cite{dem02}; 
MFT = \cite{men02}; MAG = \cite{mau04}; BTH = \cite{bat06}; 
GLM = \cite{gro09a}; HGR = \cite{hel10}.}
\end{deluxetable*}

\begin{figure} 
\includegraphics[width=2.5in]{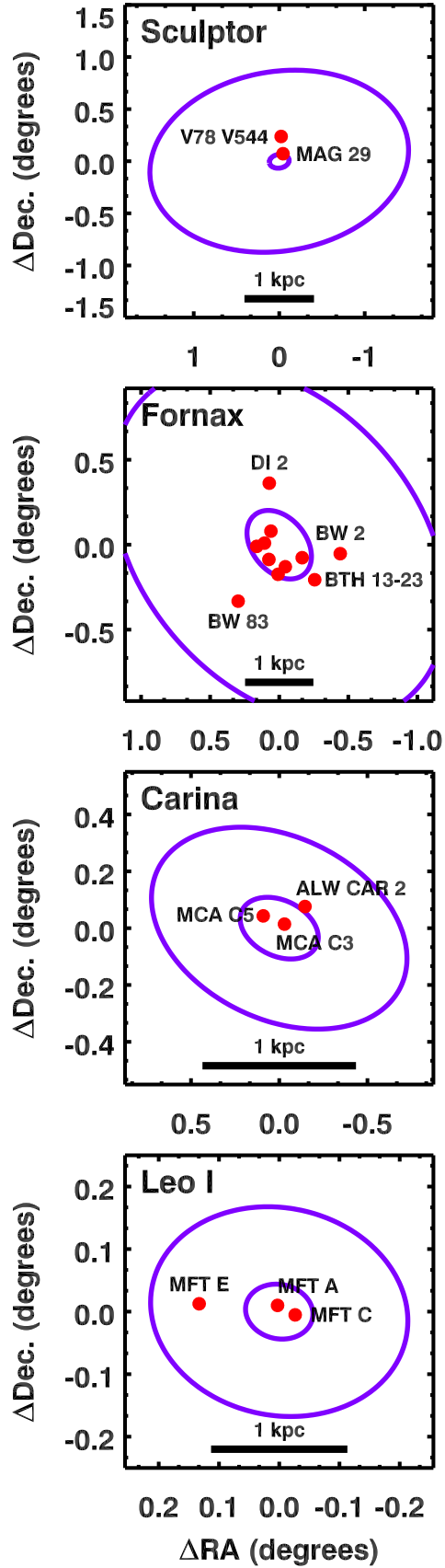}
\caption{The locations of our targets within each galaxy.  
The ellipses are the tidal and core radii defined by 
\cite{ih95}.  In Fornax, only the sources outside of the core 
are labelled.\label{f.gals}}
\end{figure}

\begin{deluxetable*}{llllll} 
\tablenum{4}
\tablecolumns{6}
\tablewidth{0pt}
\tablecaption{Optical photometry}
\label{t.phot}
\tablehead{
  \colhead{Source} & \multicolumn{4}{c}{Photometry\tablenotemark{a}} &
  \colhead{Ref.\tablenotemark{b}} \\ 
  \colhead{name} & \colhead{$B$} & \colhead{$V$} & \colhead{$R$} & 
  \colhead{$I$} & \colhead{ }
}
\startdata

MAG 29        & \nodata                 & \nodata                 &      20.22              &      18.04              & USNO-B \\
Scl V78 V544  &      20.20              &      16.55              & {\bf 16.53 $\pm$ 0.49 } &      16.70              & USNO-B \\
\\
For BW 2      & \nodata                 & {\bf 20.23} $\pm$ 0.07  & {\bf 19.21 $\pm$ 1.58 } & {\bf 16.47} $\pm$ 0.05  & BW, USNO-B \\
For BTH 13-23 & \nodata                 & \nodata                 & \nodata                 & \nodata                 & \\
For BTH 12-4  & \nodata                 & \nodata                 & \nodata                 & \nodata                 & \\
For BTH 3-129 & \nodata                 & \nodata                 & \nodata                 & \nodata                 & \\
For DK 18     &      21.45              & {\bf 18.57 $\pm$ 0.60 } & {\bf 17.51 $\pm$ 1.46 } &      16.01              & DK, USNO-B \\
For DK 52     &      22.06              & {\bf 19.74 $\pm$ 0.33 } & {\bf 17.93 $\pm$ 0.51 } &      18.15              & DK, WEL, USNO-B \\
For DI 2      &      21.4               & {\bf 18.3  $\pm$ 0.7  } & {\bf 17.07 $\pm$ 0.42 } &      16.75              & WEL, DI, USNO-B \\
For WEL C10   &      22.31 $\pm$ 0.03   & {\bf 19.41} $\pm$ 0.05  & {\bf 18.24 $\pm$ 1.08 } & {\bf 15.73} $\pm$ 0.04  & SHS, BW, USNO-B \\
For BW 69     &      21.93              & {\bf 19.99} $\pm$ 0.05  & {\bf 19.80 $\pm$ 0.74 } & {\bf 16.64} $\pm$ 0.04  & BW, USNO-B, GSC \\
For BW 75     &      20.63              & {\bf 20.06} $\pm$ 0.03  & {\bf 18.14 $\pm$ 0.33 } & {\bf 16.99} $\pm$ 0.07  & GSC, BW, USNO-B \\
For BW 83     &      23.26              & {\bf 20.41} $\pm$ 0.10  & {\bf 17.63 $\pm$ 0.98 } & {\bf 16.62} $\pm$ 0.06  & BW, USNO-B, NOMAD \\
\\
ALW Car 2     & {\bf 18.89 $\pm$ 0.64 } &      17.64              & {\bf 16.37 $\pm$ 0.01 } &      15.64              & USNO-B, NOMAD, K06 \\
Car MCA C3    &      18.43              &      17.48              &      15.86              &      15.27              & USNO-B, NOMAD \\
Car MCA C5    & {\bf 18.95 $\pm$ 0.71 } & {\bf 17.26 $\pm$ 0.09 } & {\bf 16.01 $\pm$ 0.33 } &      15.11              & M82, USNO-B, NOMAD \\
\\
Leo I MFT C   & \nodata                 & \nodata                 & \nodata                 & \nodata                 & \\
Leo I MFT A   & \nodata                 & \nodata                 & \nodata                 & \nodata                 & \\
Leo I MFT E   & \nodata                 & \nodata                 & \nodata                 & \nodata                 & 
\enddata
\tablenotetext{a}{Entries in {\bf bold} are mean magnitudes and their
  standard deviation or amplitude from multiple photometric observations.  
  If only the mean magnitude is bold, then it is followed by an uncertainty 
  in the mean.}
\tablenotetext{b}{DK = \cite{dk79}; M82 = \cite{mou82}; WEL = \cite{wes87}; 
  DI = \cite{di87}; SHS = \cite{ste98}; BW = \cite{bw02};
  USNO-B = \cite{usno03}; NOMAD = \cite{zac04}; K06 = \cite{koc06}; 
  GSC = \cite{gsc08}.}
\end{deluxetable*}

The carbon stars in our sample were observed in two {\it
Spitzer} programs.  The first, a Cycle 2 program,  included 
five carbon stars in Fornax \citep[published by][]{mat07} and 
three carbon-star candidates in Leo~I.  The second program 
followed in Cycle 3 and included six carbon stars in Fornax, 
three in Carina, and two in Sculptor.  \cite{slo09} published 
one of the two Sculptor spectra.  Table~\ref{t.star} gives 
the names, positions, and NIR fluxes from 2MASS \citep{skr06} 
of the stars in our sample.  Figure~\ref{f.gals} shows where
our targets are located in each galaxy.

Fornax is well known as an abundant source of carbon stars,
starting with the initial detection of several candidates
by \cite{dk79} and the spectroscopic confirmation of six by
\cite{am80}.  By 1999, the number of confirmed carbon stars
had climbed to 104 \citep{azz99}.  Multiple programs have 
searched Fornax for LPVs.  \cite{di87} found 30 candidates, 
but no Mira variables.  \cite{bw02} identified 85 candidates,
but did not attempt to determine periods.  \cite{whi09}
published the results of a thorough NIR monitoring program 
from the SAAO.  The Cycle 2 program selected five targets in 
Fornax based on the SAAO observations, although at that time 
the mean NIR magnitudes and periods were not known.  The six 
targets in Cycle 3 were selected based on 1--5~\mum\ spectra 
and photometry with the Infrared Spectrometer and Array 
Camera (ISAAC) on the VLT.  \cite{gro09a} published the 
results of these observations, which include eight of our 11 
targets.  The combination of the various observations 
confirms that all of our sample are carbon-rich.  The SAAO 
photometry provides periods for six of our 11 targets in 
Fornax.  Four of these are Mira variables, while two are 
semi-regulars.

Our search for targets in Sculptor began with the 2MASS 
survey.  Two targets fulfilled our criteria that $J-K_s > 
1.1$ and $K_s <$ 12.5.  V78 V544 was originally identified as 
a LPV by \cite{va78}.  This star had the reddest 
$J-K$ color (1.19) in the study by \cite{fro82}, but they were 
unable to determine if it was carbon-rich spectroscopically.  
\cite{alw86} made that confirmation, noting that their list
of eight carbon stars was likely to be complete.  However,
the optical surveys in use at that time missed the sources
embedded in the most optically thick shells.  Our other
Sculptor target, MAG 29, has $J-K_s = 3.24$.  \cite{mau04} 
first noticed this source in their search of 2MASS targets in 
the direction of several dwarf galaxies, but they estimated
its distance to be $\sim$50~kpc, in the foreground of the
Sculptor dwarf.  \cite{slo09} published an early version of
the IRS spectrum of MAG 29.  Using two different infrared
color-magnitude relations, they estimated its distance to 
be 84 $\pm$ 13 kpc, consistent with the distance of Sculptor
in Table~\ref{t.gal}, 85 $\pm$ 2 kpc.  Near-infrared spectra
of V78 V544 and MAG 29 by \cite{gro09a} confirm their 
carbon-rich nature.  \cite{men11} find that both are Mira 
variables and that the P-L relation gives distances for 
these two consistent with membership in Sculptor.

We also used the 2MASS survey to search Carina for suitable
targets.  Four sources fulfilled the criteria $J-K_s >$ 1.1
and $K_s <$ 13.0 (excluding obvious foreground sources), and 
of these we observed three\footnote{The unobserved source is 
ALW Car 7.}.  \cite{mou82} originally identified two of our
three targets as carbon stars, and we adopt their names for
them.  \cite{alw86} included all four of the red sources in 
their list of nine spectroscopically confirmed carbon stars 
in Carina.  They stated that the list should be complete for 
the area observed.

More carbon stars have been detected in Leo~I than in 
Sculptor or Carina.  \cite{alw86} listed 16 spectroscopically 
confirmed carbon stars and two candidate carbon stars in 
Leo~I.  However, none of these are very red.  A more recent 
NIR survey from the SAAO detected five highly reddened stars 
with $J-K_s >$ 2 \citep{men02}.  In our Cycle 2 program, we 
observed the three reddest, all with $J-K_s >$ 3.  
\cite{men10} determined periods for our three Leo~I targets 
as part of a larger effort which identified 26 AGB variables 
in the galaxy from the SAAO.  While it is quite likely that 
all three of our targets are carbon-rich, the spectra 
presented here are our first chance to confirm their 
chemistry.

\begin{deluxetable*}{llrlllllll} 
\tablenum{5}
\tablecolumns{10}
\tablewidth{0pt}
\tablecaption{Variability}
\label{t.var}
\tablehead{
  \colhead{Source} & \colhead{Var.} & \colhead{Period} & 
  \multicolumn{6}{c}{Photometry\tablenotemark{a}} & 
  \colhead{Ref.\tablenotemark{b}} \\
  \colhead{name} & \colhead{class} & \colhead{(days)} & 
  \colhead{$J$} & \colhead{$\Delta J$} & \colhead{$H$} & 
  \colhead{$\Delta H$} & \colhead{$K_s$} & 
  \colhead{$\Delta K_s$} & \colhead{ }
}
\startdata
MAG 29        & Mira & 554     & {\bf 14.35}            &   & {\bf 12.94}            &   & {\bf 11.44} & {\bf 0.87}   & M11 \\
Scl V78 V544  & Mira & 189     & {\bf 13.78}            &   & {\bf 12.90}            &   & {\bf 12.38} & {\bf 0.42}   & M11 \\
\\
For BW 2      & var. & \nodata &      16.05 $\pm$ 0.10  &   &      14.48 $\pm$ 0.06  &   &      13.32 $\pm$ 0.05  &   & BW, 2MASS \\
For BTH 13-23 & Mira & 350     & {\bf 17.09} & {\bf 1.28}   & {\bf 15.33} & {\bf 1.24}   & {\bf 13.63} & {\bf 1.02}   & W09 \\
For BTH 12-4  & Mira & 470     & {\bf 16.01} & {\bf 1.44}   & {\bf 14.31} & {\bf 1.14}   & {\bf 12.91} & {\bf 0.96}   & W09 \\
For BTH 3-129 & Mira & 400     & {\bf 18.00} & {\bf 1.64}   & {\bf 15.83} & {\bf 1.43}   & {\bf 13.90} & {\bf 1.16}   & W09 \\
For DK 18     & var. & \nodata & {\bf 14.71} & {\bf 1.00}   & {\bf 13.59} & {\bf 0.76}   & {\bf 12.78} & {\bf 0.43}   & W09 \\
For DK 52     & var. & \nodata & {\bf 14.76} & {\bf 0.61}   & {\bf 13.60} & {\bf 0.52}   & {\bf 12.80} & {\bf 0.32}   & W09 \\
For DI 2      & irr. & \nodata &      15.79 $\pm$ 0.07  &   &      14.56 $\pm$ 0.07  &   &      13.67 $\pm$ 0.05  &   & DI, 2MASS \\
For WEL C10   & SR   & 317     & {\bf 15.09} & {\bf 0.64}   & {\bf 13.87} & {\bf 0.43}   & {\bf 13.05} & {\bf 0.26}   & DI, W09 \\
For BW 69     & SR   & 340     & {\bf 16.24}            &   & {\bf 14.85}            &   & {\bf 13.61}            &   & W09 \\
For BW 75     & var. & \nodata & {\bf 15.12} & {\bf 1.00}   & {\bf 13.94} & {\bf 0.75}   & {\bf 13.16} & {\bf 0.51}   & W09 \\
For BW 83     & Mira & 280     & {\bf 14.87} & {\bf 0.69}   & {\bf 13.77} & {\bf 0.56}   & {\bf 12.99} & {\bf 0.52}   & W09 \\
\\
ALW Car 2     & var. & \nodata &      13.93 $\pm$ 0.03  &   &      13.07 $\pm$ 0.03  &   &      12.66 $\pm$ 0.03  &   & SAAO, 2MASS\tablenotemark{c} \\
Car MCA C3    & var. & \nodata &      13.74 $\pm$ 0.03  &   &      12.81 $\pm$ 0.02  &   &      12.34 $\pm$ 0.03  &   & SAAO, 2MASS\tablenotemark{c} \\
Car MCA C5    & var. & \nodata &      13.94 $\pm$ 0.03  &   &      13.16 $\pm$ 0.03  &   &      12.79 $\pm$ 0.03  &   & SAAO, 2MASS\tablenotemark{c} \\
\\
Leo I MFT C   & Mira & 523     & {\bf 17.46} & {\bf 1.52}   & {\bf 15.68} & {\bf 1.29}   & {\bf 13.98} & {\bf 1.03}   & M10 \\
Leo I MFT A   & Mira & 336     & {\bf 17.62} & {\bf 1.23}   & {\bf 15.88} & {\bf 1.01}   & {\bf 14.39} & {\bf 0.81}   & M10 \\
Leo I MFT E   & Mira & 283     & {\bf 19.18} & {\bf 1.87}   & {\bf 17.25} & {\bf 1.23}   & {\bf 15.63} & {\bf 1.17}   & M10
\enddata
\tablenotetext{a}{Entries in {\bf bold} are mean magnitudes and peak-to-peak
  amplitudes based on light-curve analysis.}
\tablenotetext{b}{DI = \cite{di87}; BW = \cite{bw02}; 2MASS = \cite{skr06};
  W09 =  \cite{whi09}; M10 = \cite{men10}; 
  M11 = \cite{men11}; SAAO = Unpublished communication from SAAO.}
\tablenotetext{c}{The SAAO identifies the target as variable; the photometry are from 2MASS.}
\end{deluxetable*}

Table~\ref{t.phot} gives the best available optical
photometry for the sources in our sample.  Entries in bold
are average magnitudes and the standard deviations when the
quoted sources give multiple measurements or provide a mean.
Table~\ref{t.var} presents variability classes, periods, and
NIR photometry.  The entries in bold in this table are mean 
magnitudes and peak-to-peak amplitudes published by the SAAO 
(references are given in the table notes).


\section{Observations and data reduction} 

\begin{deluxetable*}{llcccrrrr} 
\tablenum{6}
\tablecolumns{9}
\tablewidth{0pt}
\tablecaption{IRS observations}
\label{t.obs}
\tablehead{
  \colhead{Source} & \colhead{AOR} & \colhead{Program} & 
  \multicolumn{2}{c}{Observing date} &
  \multicolumn{4}{c}{On-source integration times (sec)} \\
  \colhead{name} & \colhead{key} & \colhead{ } & 
  \colhead{day} & \colhead{JD$-$2400000.5} & 
  \colhead{SL2} & \colhead{SL1} & \colhead{LL2} & \colhead{LL1}
}
\startdata

MAG 29        & 18050816 & 30333 & 2006 Dec 19 & 54088.9 &  240 &  240 &     960 &     960 \\
Scl V78 V544  & 18051072 & 30333 & 2006 Dec 20 & 54089.1 & 2880 & 5280 & \nodata & \nodata \\
\\
For BW 2      & 18053376 & 30333 & 2007 Feb 08 & 54139.7 &  480 &  480 &    2880 &    2880 \\
For BTH 13-23 & 14540544 & 20357 & 2006 Jan 30 & 53765.6 & 2880 & 3360 & \nodata & \nodata \\
For BTH 12-4  & 14541056 & 20357 & 2006 Jan 27 & 53762.9 & 2880 & 3360 & \nodata & \nodata \\
For BTH 3-129 & 14540800 & 20357 & 2006 Jan 30 & 53765.5 & 2880 & 3360 & \nodata & \nodata \\
For DK 18     & 18052864 & 30333 & 2007 Feb 08 & 54139.6 & 1440 & 1440 & \nodata & \nodata \\
For DK 52     & 18052096 & 30333 & 2006 Dec 19 & 54088.9 & 1440 & 1440 & \nodata & \nodata \\
For DI 2      & 18052352 & 30333 & 2007 Feb 06 & 54137.6 & 2880 & 2880 & \nodata & \nodata \\
For WEL C10   & 14541824 & 20357 & 2006 Jan 27 & 53762.9 & 2880 & 3360 & \nodata & \nodata \\
For BW 69     & 18052608 & 30333 & 2007 Feb 07 & 54138.9 & 1440 & 1440 & \nodata & \nodata \\
For BW 75     & 14542080 & 20357 & 2006 Jan 27 & 53762.8 & 2880 & 3360 & \nodata & \nodata \\
For BW 83     & 18053120 & 30333 & 2007 Feb 08 & 54139.7 & 1440 & 1440 & \nodata & \nodata \\
\\
ALW Car 2     & 18051584 & 30333 & 2007 Mar 12 & 54171.9 & 2880 & 5280 & \nodata & \nodata \\
Car MCA C3    & 18051328 & 30333 & 2007 Mar 12 & 54171.8 & 2880 & 4800 & \nodata & \nodata \\
Car MCA C5    & 18051840 & 30333 & 2007 Mar 13 & 54172.0 & 3840 & 5760 & \nodata & \nodata \\
\\
Leo I MFT C   & 14545152 & 20357 & 2006 May 25 & 53880.0 & 2880 & 3360 & \nodata & \nodata \\
Leo I MFT A   & 14545408 & 20357 & 2006 May 25 & 53880.1 & 2880 & 3360 & \nodata & \nodata \\
Leo I MFT E   & 14545920 & 20357 & 2006 May 25 & 53880.2 & 2880 & 3120 & \nodata & \nodata 
\enddata
\end{deluxetable*}

The IRS obtained infrared spectra of our 19 targets in the
standard staring mode, observing each target in the two
Short-Low (SL) apertures, producing spectra from 5 to 
14~\mum.  SL order 2 (SL2) covers the 5.1--7.5~\mum\ region, 
while SL order 1 (SL1) covers 7.5--14.2~\mum.  Two 
relatively bright targets were also observed in the two 
Long-Low (LL) apertures (LL2:  14--20.5~\mum; LL1:  
20.5--37~\mum).  Table~\ref{t.obs} presents the details for 
each observation.  The SL2 and LL2 spectra include a short
piece of a first-order spectrum, the ``bonus'' order.
These bonus orders provided overlap between SL2 and SL1
(and LL2 and LL1), making it possible to determine
multiplicative corrections to remove discontinuities between 
segments which arise from pointing shifts during the
IRS integrations.  Each source was observed in two separate 
nod positions in each aperture, requiring four separate 
pointings for the SL-only spectra and eight for the two which 
included LL.


Our data analysis began with the flatfielded 
images\footnote{Basic calibrated data, or BCD files.}
produced by the S18.18 pipeline at the {\it Spitzer} Science
Center (SSC).  Before extracting spectra from images, we 
removed the background by subtracting the corresponding image 
with the source in a different position, either in the other
nod position in the same aperture (a nod difference) or in
the same nod position, but in the other aperture (an aperture
difference).  The SL images in Cycle 3 used aperture
differences, but the Cycle 2 images required nod differences,
due to the mismatching number of observations in SL2 and SL1.
Nod differences were used for the two LL spectra.

The differenced IRS images were then cleaned, using the {\sc 
imclean} procedure, which is similar to the {\sc irsclean} 
procedure available from the SSC.  Pixels were replaced with 
an average computed from surrounding rows if they were 
flagged as bad or if they were included in the campaign masks 
of rogue pixels distributed by the SSC.  We generally treated 
a pixel as a rogue if it had been flagged as such twice in 
the current or any prior IRS campaign.  The number of rogue 
pixels steadily grew over the course of the cryogenic {\it 
Spitzer} mission, and not all rogue pixels were ever flagged.  
We added several additional pixels to the rogue masks for the 
Cycle 3 data when we could see the impact of consistently
misbehaving pixels on our final spectra.  This step was
crucial for improving the signal/noise (S/N) ratio of our
faint spectra, because these unflagged rogue pixels 
contribute non-gaussian noise which becomes more significant
at low flux levels.

Our extraction of spectra from the images relied on the
optimal extraction algorithm developed at Cornell and
described in detail by \cite{lev10}.  This algorithm fits a
wavelength-dependent point-spread function (PSF) to each row
of the spectral image, reducing the impact of noise from
pixels containing little flux from the source.  For point 
sources, it substantially improves the S/N compared to 
spectra extracted  with more conventional algorithms.

Spectra from individual images with the source in a given
nod position were then co-added.  We used a spike-rejection
algorithm to further reduce the effect of non-gaussian noise
components when combining the spectra from the two nod 
positions.  At that stage, we re-assessed the propagated 
noise.  If the uncertainty as measured by comparing the two
nod positions was larger, we used this value instead.
Finally, spectra from different spectral orders were 
combined using a ``stitch-and-trim'' algorithm, first
applying multiplicative corrections to remove discontinuities
between spectral segments, then truncating invalid data from
the ends of each segment.  The corrections were typically on
the order of $\sim$5\%, although they could be as large as 
15\%.

The photometric calibration of the spectra has changed 
slightly from previous publications from the IRS team at 
Cornell, as outlined by \cite{lev11}.  We use HR~6348 
(K0 III) as the standard for SL, as before, but the assumed
truth spectrum for this source has been shifted down 5\% to
align with the updated calibration of the Multiband Imaging
Photometer for {\it Spitzer} (MIPS) at 24~\mum\ 
\citep{rie08}.  We used HR~6348 and HD~173511 (K5 III) for 
LL, with the latter spectrum shifted similarly.

\begin{figure} 
\includegraphics[width=3.25in]{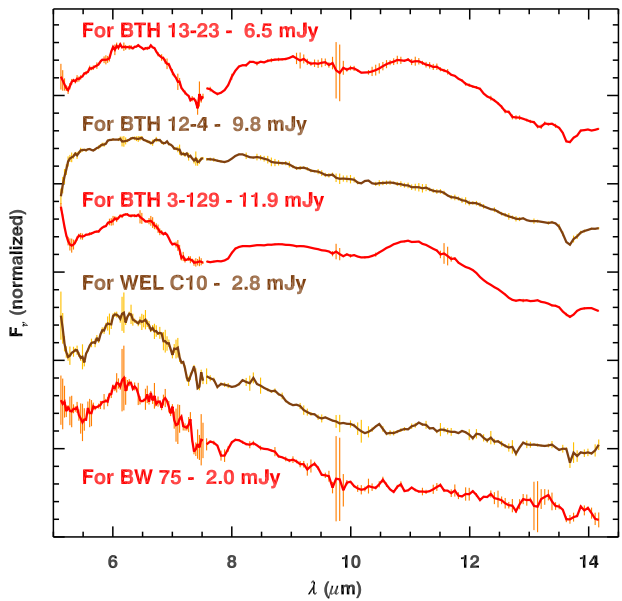}
\caption{Optimally extracted SL spectra of the five Fornax 
carbon stars originally presented by \cite{mat07}.  The 
fluxes measured at 6.0--6.5~\mum\ are given beside each 
target name.\label{f.sp1}}
\end{figure}

\begin{figure} 
\includegraphics[width=3.25in]{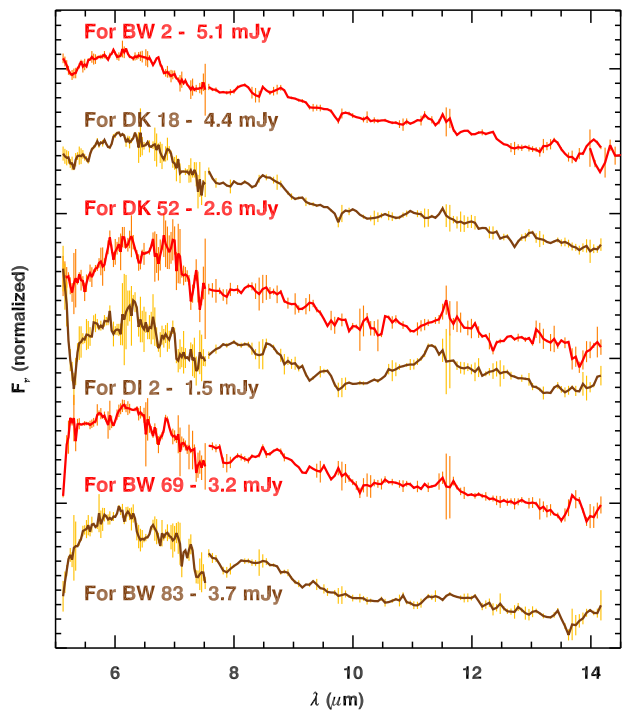}
\caption{Optimally extracted SL spectra of the six new 
carbon stars in Fornax.  Fluxes measured at 6.0--6.5~\mum\
follow each target name.\label{f.sp2}}
\end{figure}

\begin{figure} 
\includegraphics[width=3.25in]{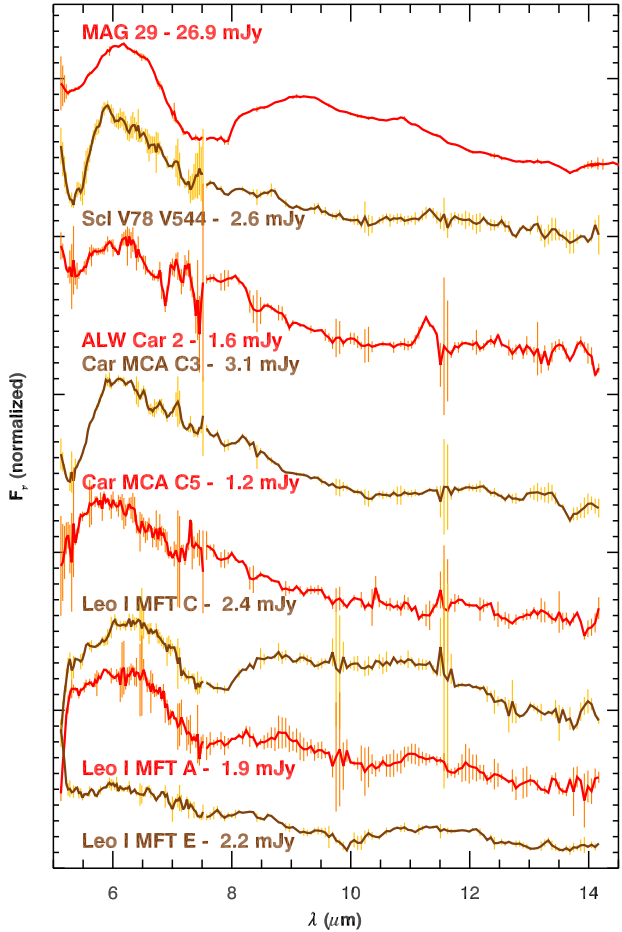}
\caption{Optimally extracted SL spectra of the sources in 
Sculptor, Carina, and Leo~I.  The fluxes after each target
name are measured at 6.0--6.5~\mum.\label{f.sp3}}
\end{figure}

\begin{figure} 
\includegraphics[width=3.25in]{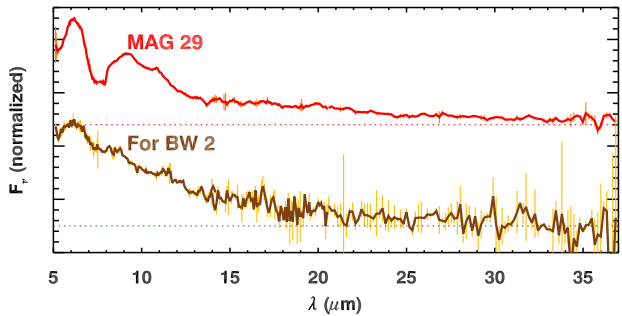}
\caption{The full IRS spectra for MAG 29 and For BW 2, which 
were observed in both SL and LL.  The dotted lines mark the
zero-flux levels for both spectra.\label{f.spll}}
\end{figure}

Figures~\ref{f.sp1}--\ref{f.sp3} present the final SL spectra
for our 19 targets.  Figure~\ref{f.spll} presents the full
spectra for the two sources also observed in LL.

\section{Refining the metallicity estimates} 

\subsection{Bolometric magnitudes} 

\begin{deluxetable*}{lccllll} 
\tablenum{7}
\tablecolumns{7}
\tablewidth{0pt}
\tablecaption{Bolometric magnitudes}
\label{t.mbol}
\tablehead{
  \colhead{Source} & \colhead{Our} & \colhead{External} &
  \multicolumn{4}{c}{Other values for $M_{\rm bol}$\tablenotemark{b}} \\
  \colhead{name} & \colhead{$M_{\rm bol}$} & 
  \colhead{uncertainty\tablenotemark{a}} &
  \colhead{M07} & \colhead{L08} & \colhead{G09} & \colhead{SAAO}
}
\startdata

MAG 29        & $-$4.90 $\pm$ 0.04 & \nodata & \nodata  & \nodata & \nodata & \nodata \\
Scl V78 V544  & $-$4.20 $\pm$ 0.04 & \nodata & \nodata  & \nodata & \nodata & \nodata \\
\\
For BW 2      & $-$4.41 $\pm$ 0.29 & 0.25    & \nodata  & \nodata & $-$4.05 & \nodata \\
For BTH 13-23 & $-$4.28 $\pm$ 0.04 & 0.30    & $-$4.90  & $-$4.84 & \nodata & $-$4.45 \\
For BTH 12-4  & $-$4.90 $\pm$ 0.04 & 0.29    & $-$5.52  & $-$5.26 & $-$5.20 & $-$4.80 \\
For BTH 3-129 & $-$4.65 $\pm$ 0.04 & 0.17    & $-$4.95  & \nodata & \nodata & $-$4.68 \\
For DK 18     & $-$4.88 $\pm$ 0.04 & 0.54    & \nodata  & \nodata & $-$4.11 & \nodata \\
For DK 52     & $-$4.62 $\pm$ 0.04 & 0.02    & \nodata  & \nodata & $-$4.65 & \nodata \\
For DI 2      & $-$4.06 $\pm$ 0.29 & 0.34    & \nodata  & \nodata & $-$3.58 & \nodata \\
For WEL C10   & $-$4.61 $\pm$ 0.04 & 0.20    & $-$4.60: & \nodata & \nodata & $-$4.21 \\
For BW 69     & $-$3.98 $\pm$ 0.04 & 0.06    & \nodata  & \nodata & $-$4.07 & \nodata \\
For BW 75     & $-$4.35 $\pm$ 0.04 & 0.22    & $-$4.67  & \nodata & $-$4.25 & \nodata \\
For BW 83     & $-$4.57 $\pm$ 0.04 & 0.19    & \nodata  & \nodata & $-$4.61 & $-$4.27 \\
\\
ALW Car 2     & $-$4.54 $\pm$ 0.29 & \nodata & \nodata  & \nodata & \nodata & \nodata \\
Car MCA C3    & $-$4.82 $\pm$ 0.29 & \nodata & \nodata  & \nodata & \nodata & \nodata \\
Car MCA C5    & $-$4.57 $\pm$ 0.29 & \nodata & \nodata  & \nodata & \nodata & \nodata \\
\\
Leo I MFT C   & $-$4.77 $\pm$ 0.13 & 0.47    & \nodata  & \nodata & \nodata & $-$5.44 \\
Leo I MFT A   & $-$4.43 $\pm$ 0.13 & 0.23    & \nodata  & \nodata & \nodata & $-$4.75 \\
Leo I MFT E   & $-$4.34 $\pm$ 0.13 & 0.42    & \nodata  & \nodata & \nodata & $-$3.74
\enddata
\tablenotetext{a}{The standard deviation of all of the given 
  values for $M_{\rm bol}$.}
\tablenotetext{b}{Adjusted to our adopted distance modulus.
  M07 = \cite{mat07}; L08 = \cite{lag08}; G09 = \cite{gro09a},
  SAAO = \cite{whi09} for Fornax and \cite{men10} for Leo~I.}
\end{deluxetable*}

We estimate bolometric magnitudes by integrating the IRS 
spectra, combined with the available optical and NIR 
photometry.  Table~\ref{t.mbol} presents the results and
compares them with previously published estimates.  Over the 
region covered by the spectrum, simple integration suffices.  
Below 5.1~\mum, we integrate on a grid of $F_{\lambda}$ vs.\ 
wavelength linearly interpolated through the photometry in 
Tables \ref{t.phot} and \ref{t.var}.  We extrapolate with a 
Rayleigh-Jeans tail at the long-wavelength end and a Wien 
distribution at the short-wavelength end.  The final 
magnitudes are scaled to the distances given in 
Table~\ref{t.gal}.

This paper marks a shift from previous determinations of
bolometric magnitude.  Before, we did not consider $BVRI$ 
photometry.  The change is significant.  The Wien
distribution extrapolated from $J$ overestimates the 
flux in the visual regime by $\sim$0.2 magnitudes for our 
bluest sources, because the combination of molecular band 
absorption and dust extinction drops more quickly with 
decreasing wavelength.  The largest difference in our sample 
is 0.27 magnitudes\footnote{The star was For~DK~52.}.  The 
difference decreases to zero past $J-K \sim 3$.  A line can 
be fitted to this shift:  $\Delta M_{\rm bol} = 0.27 - 0.075 
(J-K)$, although the scatter is $\sim$0.1 magnitudes.  It 
would be appropriate to apply this correction to our 
previously published bolometric magnitudes for carbon stars 
\citep[e.g.,][and most of the other IRS papers cited in 
\S 1]{slo06}.

We have simplified the treatment of the $BVRI$ photometry by
not considering differences in photometric systems or 
attempting transformations among them.  Such an effort would
have virtually no impact on the bolometric magnitudes
reported here.  However, for the NIR photometry, we did
distinguish between the SAAO and 2MASS systems, as it is in
this wavelength regime that the spectral energy distributions
(SEDs) of our sources peak\footnote{\cite{coh03} defined the
the central wavelengths and zero-magnitude fluxes for 2MASS.
For SAAO, \cite{nag03} defined the wavelengths, and we scaled
the zero-magnitude fluxes from the 2MASS data.}.


The impact of the correction for interstellar reddening is
much smaller.  It brightens the bluest stars by only 
$\sim$0.02 magnitudes in our samples, with the correction 
decreasing to nearly zero for the most enshrouded carbon 
stars.  The redder the $J-K$ color, the more the peak of 
the SED has shifted away from wavelengths most affected by 
interstellar reddening.

The most significant source of uncertainty in our sample is
the variability of the star.  The $JHK$ photometry dominates
the final bolometric magnitude.  Comparing the mean
magnitudes from the SAAO to the 2MASS data reveals 
an average difference ($|\Delta M_{\rm bol}|$) of 0.29 
magnitudes, in either direction, with differences as large as
0.6 magnitudes.  Interestingly, we get similarly large 
differences if we only consider the sources identified as 
semi-regulars, irregulars, or simply ``variables''.  For any 
variable star, the mean magnitude is definitely preferable to 
NIR photometry that only sparsely covers the period of 
variability.

For the sources with mean NIR magnitudes, the uncertainties 
in bolometric magnitude in Table~\ref{t.mbol} are just the
uncertainty in distance modulus.  For the remaining five 
sources, we set the uncertainties to 0.29 to reflect the 
limitations of the 2MASS photometry.  The reader should bear 
in mind, though, that errors as large as 0.6 magnitudes are 
possible.

Table~\ref{t.mbol} also lists what we describe as 
``external'' uncertainties in $M_{\rm bol}$.  These are the 
standard deviation of all values of $M_{\rm bol}$ for a given 
star in Table~\ref{t.mbol}.  The values from other authors 
have been adjusted to our adopted distance moduli (in 
Table~\ref{t.gal}).  The values published by \cite{mat07} are
based on earlier versions of the IRS data for the five stars
in common between this work and theirs.  They used these data
along with the NIR photometry from the SAAO and fitted 
radiative transfer models to determine the luminosity.  Four 
of their bolometric magnitudes appear in their Table~4; we 
reconstructed the fifth from the luminosity given in the 
text.  \cite{lag08} used 2MASS photometry and applied the
bolometric corrections defined by \cite{whi06}, which were
calibrated from Galactic carbon stars with photometry from
the optical into the mid-infrared, including mean magnitudes
at $JHK$, {\it IRAS} data to 25~\mum\ and {\it MSX} data to 
15~\mum\footnote{{\it IRAS:}  the {\it Infrared Astronomical 
Satellite} \citep{bei88}; {\it MSX:}  the {\it Mid-course 
Space Experiment} \citep{ega03}.}.  \cite{gro09a} applied the 
bolometric corrections defined by \cite{bw84} to 2MASS 
photometry.  The bolometric magnitudes from the SAAO in 
Table~\ref{t.mbol} are based on mean magnitudes at $JHK$ and
the bolometric corrections of \cite{whi06}.  The external 
uncertainties may well overestimate our actual uncertainty in 
$M_{\rm bol}$ as they reflect systematic differences between 
groups with access to different data, but including these 
systematics will force us to be cautious with our $M_{\rm bol}$ 
results.

\subsection{Masses and metallicities\label{s.mm}} 

\begin{figure} 
\includegraphics[width=3.25in]{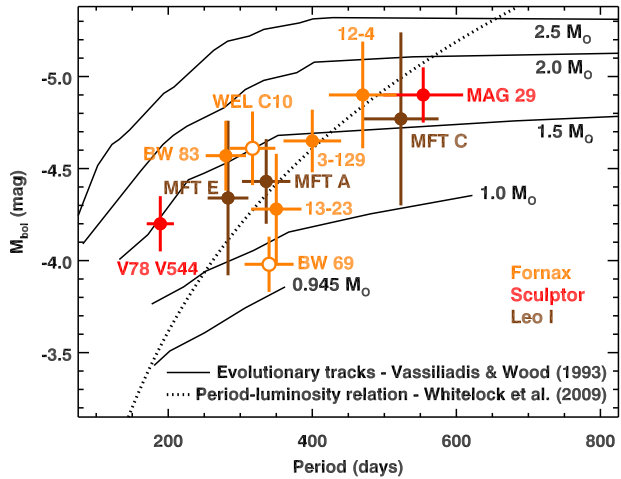}
\caption{The 11 sources for which we have periods plotted
on a period-luminosity (P-L) diagram and compared to 
evolutionary tracks by \cite{vw93} and the P-L relation of
\cite{whi09}.  The masses given are initial values in solar
masses.  For $M_{\rm bol}$, the error bars are the larger 
of the uncertainties in Table~\ref{t.mbol}, with a minimum 
assumed uncertainty of 0.15 magnitudes (which affects Fornax 
BW 69 and the two Sculptor objects).  We have assumed a 10\% 
uncertainty in period.  To avoid clutter in the figure, we 
have shortened the source names.  The two open circles mark 
the semi-regular variables.\label{f.evo}}
\end{figure}

\begin{deluxetable}{lcc} 
\tablenum{8}
\tablecolumns{3}
\tablewidth{0pt}
\tablecaption{Possible stellar ages}
\label{t.age}
\tablehead{
  \colhead{Source} & \colhead{Estimated} & \colhead{Estimated} \\
  \colhead{name}   & \colhead{initial mass ($M_{\odot}$)} & \colhead{age (Gyr)} 
}
\startdata

MAG 29        &  1.5--1.9 & 0.9--1.7 \\
Scl V78 V544  &  1.2--1.8 & 1.0--3.5 \\
\\
For BTH 13-23 & 0.97--1.5 & 1.8--11 \\
For BTH 12-4  &  1.4--2.2 & 0.6--2.7 \\
For BTH 3-129 &  1.3--1.7 & 1.2--3.3 \\
For WEL C10   &  1.3--1.9 & 0.9--3.3 \\
For BW 69     & 0.94--1.1 & 5.5--12 \\
For BW 83     &  1.3--2.0 & 0.8--3.3 \\
\\
Leo I MFT C   &  1.0--2.3 & 0.5--6.9 \\
Leo I MFT A   &  1.0--1.5 & 1.7--6.9 \\
Leo I MFT E   & 0.98--1.9 & 0.8--7.3
\enddata
\end{deluxetable}

We have pulsation periods for 11 of our 19 targets, 
including both carbon stars in Sculptor, all three in Leo~I, 
and six of the 11 in Fornax (see Table~\ref{t.var}).
Figure~\ref{f.evo} compares their periods and bolometric 
magnitudes to evolutionary tracks by \cite{vw93}, allowing us 
to make rough estimates of their initial mass, and thus their 
age.  The evolutionary tracks convert the relatively robust
period-mass-radius relation from stellar pulsation theory 
into a period-mass-luminosity relation by assuming a relation
between effective temperature and luminosity.  The mass-loss
history must be estimated to convert current to initial mass,
and the histories utilized by \cite{vw93} have successfully 
reproduced the luminosities at the tips of the 
RGB in Magellanic clusters.  More theoretical approaches to
the problem \citep[e.g.,][]{kam11} basically validate the 
older evolutionary tracks.

The error bars in Figure~\ref{f.evo} are the larger of the 
internal and external uncertainties for $M_{\rm bol}$ in 
Table~\ref{t.mbol}, although we have assumed a minimum 
uncertainty of 0.15 magnitudes, which affects MAG~29 and 
V78~V544 in Sculptor and For~BW~69.  For period, we assumed a 
10\% uncertainty, based on the fact that \cite{gro09b} have
found shifts in period on this order for some of the sources
in their sample (not counting those sources with possible
changes in pulsation mode).  Our conservative approach to the 
uncertainties leads to a fairly wide range of initial masses 
and ages in our sample, but even these broad limits can help
constrain the metallicities.

Table~\ref{t.age} gives our rough estimates of initial masses 
and current ages, based on Figure~\ref{f.evo}.  Assuming that
the stellar lifetimes scale as $M^{-2.5}$ would overestimate
the ages, because more metal-poor stars evolve off of the
main sequence more quickly.  To estimate the ages, we have
used models by \cite{mar05}, who give main-sequence turn-off
masses at metallicities of [Fe/H] = $-$1.35, $-$0.33, and 
0.0.  The first metallicity matches Leo I perfectly.  For 
Sculptor we spline interpolated to [Fe/H] = $-$1.0.  For 
Fornax, we spline interpolated to $-$0.3 and $-$0.8, using 
the former for the lower bound on the mass and the latter for 
the upper bound.  To get from main-sequence turn-off point to 
the AGB, we assumed that a star spends a time on the red 
giant branch equal to 10\% of its main-sequence lifetime.

The likely ages of both stars in Sculptor are consistent with 
the models of \cite{rev09}, which show virtually no star 
formation in Sculptor over the period from $\sim$8 to 
$\sim$2~Gyr ago.  Thus both stars must be younger than 
$\sim$2~Gyr old.  Their models indicate that stars produced 
in this recent round of star formation have [Fe/H] $\sim$ 
$-$1.0.  This value is significantly higher than the mean in 
Table~\ref{t.gal} and the value of $-$1.4 assumed by 
\cite{slo09}.


\cite{men11} also placed MAG~29 in the younger population in 
Sculptor, with an age of $\sim$1--2~Gyr.  However, they made 
this estimate based on its pulsation period, which they argue 
is a good diagnostic for age (see their \S~6 and references 
therein.  They assigned Scl~V78~V544 in the older population
due to its shorter pulsation period, while we find that its 
luminosity is more consistent with the younger population 
(age $\la$ 2~Gyr).  The discrepancy between the two 
approaches arises from the significant spread in periods 
possible for stars of a given mass.

Two of the targets in Fornax, WEL~C10 and BW~69, are
semi-regular pulsators.  The location of For~BW~69 
amongst the other data suggests that it is a fundamental-mode 
pulsator.  If it were pulsating in the first overtone, then
its position in Figure~\ref{f.evo} would correspond to a
period $\sim$2.2 times longer \citep{ws96}, which would imply 
an unreasonably low mass.  If For~WEL~C10 were pulsating in 
the first overtone, it would shift to the right-most point in
the figure and become something of an outlier.  While we 
suspect that it is pulsating in the fundamental mode, an
error here would have only a small impact on its estimated 
mass, because the track for 1.5~M$_{\odot}$ is nearly 
horizontal. 

Four of the six sources with periods in Fornax have relatively
young ages of $\sim$3.3 Gyr or less.  The models of 
\cite{rev09} suggest stars of this age would have 
corresponding metallicities in the range $-$0.8 $<$ [Fe/H] 
$<$ $-$0.3.  This range is fairly constant over the likely 
time frame, and it is more similar to the Magellanic samples 
than to the other dwarf spheroidals considered here.  The
uncertain mass of BTH~13-23 leads to an unconstrained 
age and a poorly constrained metallicity.  BW~69 looks to be
at least 5.5~Gyr old.  Using the models of \cite{rev09} as a
guide, its maximum metallicity is $\sim$$-$0.5, the mean
metallicity of the SMC, although it cold be more metal-poor.

Five stars in Fornax without periods do not appear in
Figure~\ref{f.evo}.  Two were outside the survey area of
\cite{whi09}, and both were identified as variables by
other observers (see Table~\ref{t.var}).  \cite{whi09}
identified the other three as variables but were unable to
report a period.  All five bolometric magnitudes do little
to constrain their likely masses, ages, and metallicities,
which we will assume are similar to the sources for which 
we have periods.


The three carbon stars in Leo~I are probably younger than
$\sim$7~M$_{\odot}$, but their relatively unconstrained
masses allow us to say little more about their age.
Fortunately for our efforts to constrain their metallicity,
the MDF for Leo~I is narrow ($<$[Fe/H]$>$ = $-$1.35 $\pm$ 
0.24, see \S~\ref{s.metal}).  No matter their mass and age,
these stars are likely to be more metal-poor than those in
Fornax and even Sculptor.

The SAAO reports that all three of our targets in Carina are 
semi-regular variables, but the periods are undetermined (J.\ 
Menzies, P.\ Whitelock, and M.~W.\ Feast, 2011, private 
communication).  
As with Leo~I, the metallicities in Carina
are distributed more narrowly than in Sculptor or Fornax:
$<$[Fe/H]$>$ = $-$1.73 $\pm$ 0.35, making it likely that
these targets are even more metal-poor than those in Leo~I.
The metallicity measurement by \cite{abi08} reinforces this
point; they found [M/H] = $-$1.7 for MCA C3\footnote{And 
[M/H] = $-$1.9 for ALW Car 7.}.

Thus the metallicities of the carbon stars observed in Fornax 
and Sculptor are probably higher than previously proposed.  
While \cite{mat07} adopted [Fe/H] $\sim$ $-$1.0 for the five 
Fornax stars they examined, we find that [Fe/H] appears to be 
more Magellanic in nature: $\sim$$-$0.3 to $-$0.8.  
Similarly, while \cite{slo09} suggested that MAG~29, which is 
also associated with significant quantities of carbon-rich 
dust, would have formed with [Fe/H] $\sim$ $-$1.4, we have 
revised this metallicity up to $\sim$$-$1.0.

Figure~\ref{f.evo} includes the P-L relation of \cite{whi09}: 
$M_{\rm bol}$ = $-$3.3 log $P$ (days) + 3.979.  This relation 
appears on the plot as a narrow dotted curve, but for typical
data, it is $\sim$0.4--0.5 magnitudes wide at any period.  
Most of our data fall within this range.  The evolutionary 
tracks show why a significant spread in periods would be 
expected for stars of a given mass (and therefore age).  For 
stars of mass $\ga$ 1.5~M$_{\odot}$, once they reach 
pulsation periods of $\sim$350--400~days, their period will 
continue to increase, but their luminosity will remain 
largely fixed.  Thus the width of the P-L relation depends on 
how long AGB stars will survive once they begin pulsating in 
the fundamental mode.  This width can lead to considerable
uncertainty in any distances determined using P-L relations 
for LPVs, and for this reason we did not include distances 
based on the periods of LPVs in \S~\ref{s.gal} above.


\subsection{Comparison samples} 

In the analysis below, we compare the carbon stars we 
have observed in the four targeted dwarf spheroidal galaxies
with similar infrared spectroscopy of samples in the Milky
Way, LMC, and SMC.  The Galactic spectra are from the atlas
of spectra from the Short-Wavelength Spectrometer (SWS) on 
the {\it Infrared Space Observatory} ({\it ISO}) 
\citep{slo03}.  The sample includes the 37 sources classified 
by \cite{kra02} as carbon stars, including nine stars 
observed multiple times.  Using the multiple observations, we 
have arrived at typical variations over the pulsation cycle 
of the star in the strengths of the extracted features; these 
are plotted in the relevant figures to give an idea of the 
systematic uncertainties in the data.  See the description by 
\cite{slo06} for more detail on the Galactic sample.  A paper 
in preparation will present the spectral properties of this 
sample in more detail.

The Magellanic samples of carbon stars come from the 
following programs:  200, 1094, 3277, 3426, 3505, and 3591.
The publications arising from these programs are referenced 
in \S 1.  These samples include a total of 72 carbon stars in 
the LMC and 34 in the SMC.  

Some caution is required when comparing spectral data amongst 
the samples, as a variety of selection criteria bias them in 
different ways.  For example, Programs 1094 and 3591 focused 
on dusty sources in the LMC, thus selecting against optically 
thin dust shells, while Program 3426 sampled the brightest 
infrared sources in the LMC, which again selected against 
optically thin dust shells.  The impact of these particular 
criteria is readily apparent in the following figures.  For
this reason, we will attempt to compare the data in the
different samples by plotting the variable of interest 
against a dependent variable.  The effect of metallicity will 
reveal itself through changes in the dependency from one 
galaxy to the next.

\section{Spectral analysis} 

\subsection{The Manchester Method\label{s.method}} 

\begin{figure} 
\includegraphics[width=3.25in]{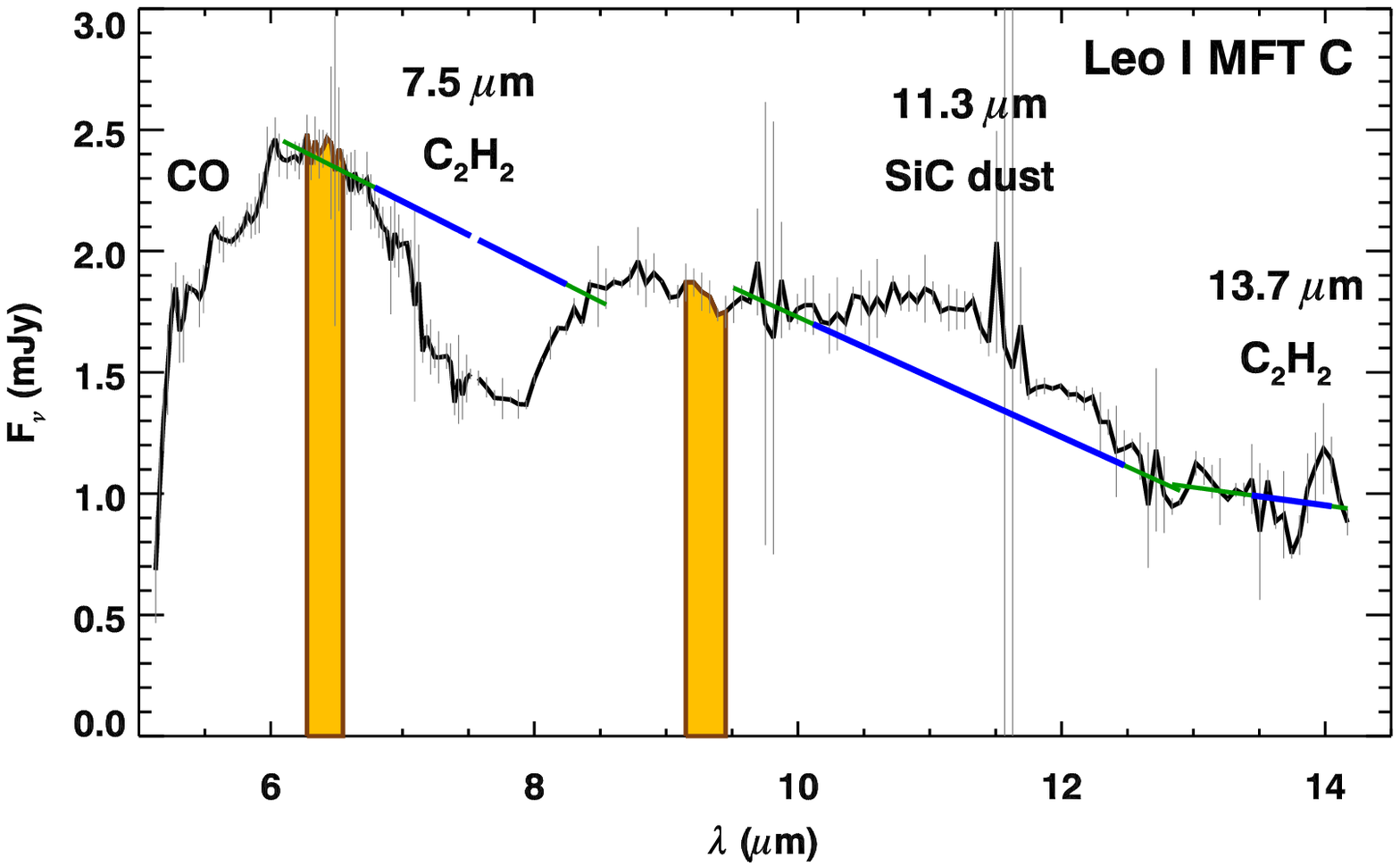}
\caption{An example of the carbon-rich dust and molecular 
features in the spectrum of Leo~I~MFT~C.  We use the 
Manchester Method to measure the [6.4]$-$[9.3] color, the 
strength of the SiC dust emission feature at 11.3~\mum\ (or 
other dust features in the vicinity), and the acetylene 
absorption bands at 7.5 and 13.7~\mum.  Note that the latter 
is just the Q branch of a much broader feature.  The method
estimates the strength of the SiC and C$_2$H$_2$ features
line segments, with the continua estimated over the ranges
where the lines are thin and the features measured where
the lines are thick.
Table~\ref{t.dust} gives the [6.4]$-$[9.3] colors and SiC
feature strengths, while Table~\ref{t.gas} gives the
equivalent widths for the acetylene bands.\label{f.mm}}
\end{figure}

\begin{deluxetable}{lccc} 
\tablecolumns{4}
\tablewidth{0pt}
\tablenum{9}
\tablecaption{Fitting wavelengths}
\label{t.mm}
\tablehead{
  \colhead{Feature} & \colhead{$\lambda$} &
  \colhead{Blue continuum} & \colhead{Red continuum} \\
  \colhead{} & \colhead{(\mum)} & \colhead{(\mum)} & \colhead{(\mum)} }
\startdata
C$_2$H$_2$ abs.         &   7.5  &  6.08--6.77  &  8.25--8.55 \\
SiC dust em.            &  11.3  &  9.50--10.10 & 12.50--12.90\\
C$_2$H$_2$ abs.         &  13.7  & 12.80--13.40 & 14.10--14.70\tablenotemark{a}
\enddata
\tablenotetext{a}{14.10--14.17 for all but the two spectra with LL data.}
\end{deluxetable}

\begin{deluxetable*}{lrcccr} 
\tablenum{10}
\tablecolumns{6}
\tablewidth{0pt}
\tablecaption{Spectral dust properties}
\label{t.dust}
\tablehead{
  \colhead{Source} & \colhead{[6.4.]$-$[9.3]} & 
    \multicolumn{2}{c}{Mass-loss rates (M$_{\odot}$/yr) } &
    \multicolumn{2}{c}{``SiC'' feature strength} \\
  \colhead{name} & \colhead{(mags.)} &  
    \colhead{total ($\log \dot{M}$)\tablenotemark{a}} & 
    \colhead{dust ($\log \dot{D}$)\tablenotemark{b}} &
    \colhead{$\lambda_c$ (\mum)} & \colhead{feature/continuum} 
}
\startdata
MAG 29        &    0.432 $\pm$ 0.008 & $-$5.91 & $-$8.21 & 10.87 $\pm$ 0.08 &    0.028 $\pm$ 0.005 \\
Scl V78 V544  & $-$0.147 $\pm$ 0.016 & \nodata & \nodata & \nodata          &    0.061 $\pm$ 0.034 \\
\\
For BW 2      &    0.333 $\pm$ 0.015 & $-$6.07 & $-$8.37 & 11.51 $\pm$ 0.16 &    0.084 $\pm$ 0.013 \\
For BTH 13-23 &    0.716 $\pm$ 0.006 & $-$5.45 & $-$7.75 & 11.29 $\pm$ 0.07 &    0.133 $\pm$ 0.006 \\
For BTH 12-4  &    0.602 $\pm$ 0.005 & $-$5.64 & $-$7.94 & 11.27 $\pm$ 0.12 &    0.045 $\pm$ 0.004 \\
For BTH 3-129 &    0.592 $\pm$ 0.006 & $-$5.65 & $-$7.95 & 11.23 $\pm$ 0.04 &    0.173 $\pm$ 0.005 \\
For DK 18     &    0.152 $\pm$ 0.021 & $-$6.36 & $-$8.66 & 11.48 $\pm$ 0.28 &    0.120 $\pm$ 0.024 \\
For DK 52     &    0.198 $\pm$ 0.031 & $-$6.28 & $-$8.58 & 11.50 $\pm$ 0.43 &    0.093 $\pm$ 0.031 \\
For DI 2      & $-$0.153 $\pm$ 0.068 & \nodata & \nodata & 11.34 $\pm$ 0.15 &    0.317 $\pm$ 0.053 \\
For WEL C10   & $-$0.046 $\pm$ 0.022 & \nodata & \nodata & 11.80 $\pm$ 0.33 &    0.051 $\pm$ 0.015 \\
For BW 69     &    0.232 $\pm$ 0.023 & $-$6.23 & $-$8.53 & \nodata          &    0.011 $\pm$ 0.026 \\
For BW 75     &    0.156 $\pm$ 0.019 & $-$6.35 & $-$8.65 & \nodata          &    0.032 $\pm$ 0.018 \\
For BW 83     &    0.163 $\pm$ 0.032 & $-$6.34 & $-$8.64 & 11.93 $\pm$ 0.26 &    0.041 $\pm$ 0.014 \\
\\
ALW Car 2     & $-$0.025 $\pm$ 0.028 & \nodata & \nodata & \nodata          &    0.012 $\pm$ 0.047 \\
Car MCA C3    & $-$0.157 $\pm$ 0.018 & \nodata & \nodata & \nodata          & $-$0.008 $\pm$ 0.039 \\
Car MCA C5    & $-$0.031 $\pm$ 0.031 & \nodata & \nodata & \nodata          &    0.088 $\pm$ 0.056 \\
\\
Leo I MFT C   &    0.500 $\pm$ 0.015 & $-$5.80 & $-$8.10 & 11.24 $\pm$ 0.45 &    0.155 $\pm$ 0.043 \\
Leo I MFT A   &    0.069 $\pm$ 0.016 & $-$6.49 & $-$8.79 & \nodata          &    0.080 $\pm$ 0.038 \\
Leo I MFT E   &    0.226 $\pm$ 0.012 & $-$6.24 & $-$8.54 & 11.32 $\pm$ 0.22 &    0.196 $\pm$ 0.026 \\
\enddata
\tablenotetext{a}{Assuming a gas-to-dust ratio ($\psi$) of 200.}
\tablenotetext{b}{Assuming an outflow velocity ($v_{\rm out}$) of 10 km/s.}
\end{deluxetable*}

\begin{deluxetable*}{lcrcr} 
\tablenum{11}
\tablecolumns{5}
\tablewidth{0pt}
\tablecaption{Spectral acetylene properties}
\label{t.gas}
\tablehead{
  \colhead{Source} & \multicolumn{2}{c}{7.5~\mum\ C$_2$H$_2$ band} &
    \multicolumn{2}{c}{13.7~\mum\ C$_2$H$_2$ band (Q branch)} \\
  \colhead{name} &  \colhead{$\lambda_c$ (\mum)} & 
  \colhead{EW (\mum)} & \colhead{$\lambda_c$ (\mum)} & 
  \colhead{EW (\mum)}
}
\startdata

MAG 29        & 7.49 $\pm$ 0.02 &    0.780 $\pm$ 0.012 & 13.69 $\pm$ 0.04 &    0.079 $\pm$ 0.007 \\
Scl V78 V544  & 7.31 $\pm$ 0.15 &    0.186 $\pm$ 0.013 & \nodata          &    0.144 $\pm$ 0.031 \\
\\
For BW 2      & 7.34 $\pm$ 0.09 &    0.112 $\pm$ 0.010 & \nodata          & $-$0.002 $\pm$ 0.029 \\
For BTH 13-23 & 7.43 $\pm$ 0.02 &    0.238 $\pm$ 0.004 & 13.68 $\pm$ 0.03 &    0.042 $\pm$ 0.004 \\
For BTH 12-4  & 7.52 $\pm$ 0.08 &    0.050 $\pm$ 0.004 & 13.70 $\pm$ 0.02 &    0.058 $\pm$ 0.003 \\
For BTH 3-129 & 7.42 $\pm$ 0.02 &    0.191 $\pm$ 0.005 & 13.66 $\pm$ 0.02 &    0.030 $\pm$ 0.002 \\
For DK 18     & 7.46 $\pm$ 0.11 &    0.160 $\pm$ 0.011 & \nodata          &    0.040 $\pm$ 0.032 \\
For DK 52     & \nodata         &    0.117 $\pm$ 0.024 & 13.80 $\pm$ 0.08 &    0.157 $\pm$ 0.027 \\
For DI 2      & 7.24 $\pm$ 0.13 &    0.183 $\pm$ 0.035 & \nodata          &    0.122 $\pm$ 0.051 \\
For WEL C10   & 7.47 $\pm$ 0.05 &    0.178 $\pm$ 0.008 & \nodata          &    0.095 $\pm$ 0.013 \\
For BW 69     & 7.44 $\pm$ 0.15 &    0.149 $\pm$ 0.021 & \nodata          &    0.034 $\pm$ 0.021 \\
For BW 75     & 7.52 $\pm$ 0.11 &    0.136 $\pm$ 0.009 & \nodata          &    0.019 $\pm$ 0.021 \\
For BW 83     & 7.48 $\pm$ 0.16 &    0.146 $\pm$ 0.020 & 13.66 $\pm$ 0.09 &    0.130 $\pm$ 0.023 \\
\\
ALW Car 2     & \nodata         & $-$0.074 $\pm$ 0.023 & \nodata          & $-$0.341 $\pm$ 0.034 \\
Car MCA C3    & \nodata         &    0.040 $\pm$ 0.014 & 13.70 $\pm$ 0.09 &    0.150 $\pm$ 0.020 \\
Car MCA C5    & \nodata         & $-$0.018 $\pm$ 0.018 & \nodata          &    0.127 $\pm$ 0.028 \\
\\
Leo I MFT C   & 7.52 $\pm$ 0.05 &    0.315 $\pm$ 0.010 & \nodata          & $-$0.008 $\pm$ 0.020 \\
Leo I MFT A   & 7.43 $\pm$ 0.08 &    0.251 $\pm$ 0.016 & \nodata          &    0.083 $\pm$ 0.036 \\
Leo I MFT E   & 7.53 $\pm$ 0.14 &    0.075 $\pm$ 0.013 & \nodata          &    0.039 $\pm$ 0.019
\enddata
\end{deluxetable*}

The mid-infrared spectra of carbon stars are rich in emission
and absorption features.  Figure~\ref{f.mm} illustrates how
we sort these out for one spectrum, using the Manchester 
Method, which was first developed for IRS spectra of carbon
stars in the Magellanic Clouds \citep{slo06,zij06}.  The 
[6.4]$-$[9.3] color samples the spectra at 6.25--6.55 and 
9.10--9.50~\mum.  These two wavelength intervals are 
relatively free of emission or absorption features, and as
detailed in \S~\ref{s.dust} below, can be used to estimate
the total amount of dust emission in the spectrum.  For ease 
of reference, we will refer to this quantity as ``dust 
content'' in the following discussion.  To measure the 
relative strengths of the acetylene bands and SiC emission 
features in the spectra, we use line segments to estimate the 
continuum above or below the feature.  Table~\ref{t.mm} gives 
the wavelengths used to fit continua and measure the various 
features.  For the molecular bands, we report an equivalent
width (EW); for the SiC dust emission, we report its total 
integrated strength, divided by the continuum underneath, as
estimated by the fitted line segment.  For all features, we 
also report a central wavelength, defined as the wavelength 
which bisects the integrated flux of the feature.  The
uncertainty in the central wavelength indicates the range
possible given the uncertainty in the extracted strength.

Table~\ref{t.dust} presents the [6.4]$-$[9.3] color and the 
relative strength of the SiC dust emission feature at 
$\sim$11.3~\mum.  The SiC feature is covered below in 
\S~\ref{s.sic}.  Table~\ref{t.gas} presents the equivalent 
widths and central wavelengths of the absorption bands from 
acetylene gas, as described in \S~\ref{s.gas}.

\subsection{Dust content and pulsation period\label{s.dust}} 

\begin{figure} 
\includegraphics[width=3.25in]{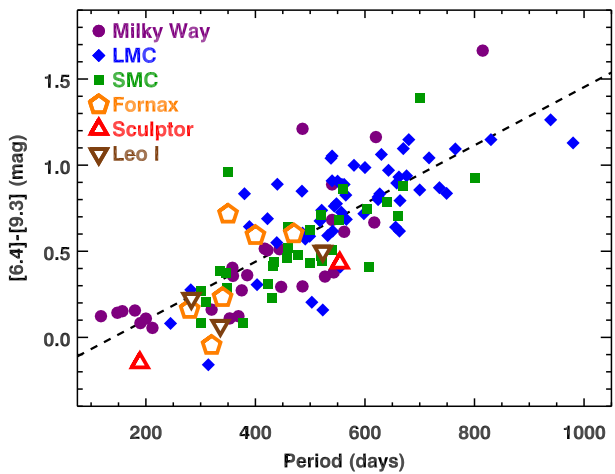}
\caption{The relation between [6.4]$-$[9.3] color and 
pulsation period for our sample of carbon stars in Local
Group dwarfs, compared to samples from the Galaxy, LMC, and 
SMC.  The uncertainties in [6.4]$-$[9.3] color are generally
smaller than the plotting symbols, and it is reasonable to
assume that the uncertainty in period is about 10\%.  The 
dashed curve is a line fitted to all of the stars with 
periods in the Milky Way, LMC, SMC, and Fornax.  All four 
galaxies show a similar dependence of total dust content on 
pulsation period, despite their differences in metallicity.  
However, all five stars in Sculptor and Leo~I lie below the 
line at a statistically significant level, indicating that 
the impact of metallicity is revealing itself for the most 
metal-poor stars sampled.\label{f.p_69}}
\end{figure}

The [6.4]$-$[9.3] color measures the total emission from the 
star plus shell in two wavelength ranges between the various
absorption and emission features.  Amorphous carbon, which 
dominates the dust around carbon stars \cite[see][]{mr87}, 
has no resonances in the infrared, so that the apparent 
``continuum'' in a spectrum is actually a combination of 
star plus dust.  The [6.4]$-$[9.3] color provides a means of 
measuring the relative combinations of these two components
and thus serves as a proxy for total dust content.

\citet[][their Fig.\ 7]{gro07} calibrated the [6.4]$-$[9.3]
color to total mass-loss rate by applying radiative transfer 
models to a large sample of IRS spectra of Magellanic carbon 
stars.  They found a linear relationship between the log of
the total mass-loss rate and the [6.4]$-$[9.3] color.  
Because their models used the same gas-to-dust ratio ($\psi$
= 200) and the same outflow velocity ($v_{\rm out}$ = 10 km/s)
for all stars, their calibration of the [6.4]$-$[9.3] color
actually ties it directly to the total mass of warm dust
contributing in the 6--10~\mum\ spectral region.
Dividing $\dot{M}$ by the gas-to-dust ratio gives the dust 
production rate ($\dot{D}$, DPR, or dust MLR), and dividing
$\dot{D}$ by the outflow velocity gives a quantity 
proportional to the optical depth of the radiating dust, which
we are calling the dust content.

\cite{slo08} presented the dust-color relation in terms of 
log $\dot{D}$ vs.\ [6.4]$-$[9.3].  Adding a term to account 
for the outflow velocity and correcting their typographical 
error,
\begin{eqnarray}
  \nonumber \log~\dot{D} \left( \frac{{\rm M_{\odot}}}{{\rm yr}} \right) = 
  \log \, \left( \frac{v_{{\rm out}}}{10 \, {\rm km/s}} \right) \\
  \times \, \{ ~ -8.9 + 1.6 \, ([6.4]-[9.3]) ~ \} .
\end{eqnarray}
This equation makes no assumptions about outflow velocity,
and it is free of any dependence on gas-to-dust ratio.
We will return to the question of how these quantities vary 
with metallicity in \S~\ref{s.vout} below.

Table~\ref{t.dust} gives the [6.4]$-$[9.3] color for each 
source.  In order to translate these colors into more 
familiar quantities, Table~\ref{t.dust} also provides 
$\dot{D}$ using the above equation and an assumed outflow 
velocity of 10 km/s and $\dot{M}$ assuming $\psi$ = 200.  It 
is important to remember, though, that the [6.4]$-$[9.3] 
color actually measures the dust content.

\cite{slo08} compared the dust content in carbon stars in
Magellanic and Galactic samples by plotting the [6.4]$-$[9.3] 
color as a function of the pulsation period of the star 
(their Fig.~29).  They found that the amount of dust as 
measured by the [6.4]$-$[9.3] color increases with pulsation 
period.  The scatter in [6.4]$-$[9.3] color at a given 
period is substantial.  Within this envelope, no
dependency on metallicity was apparent between the samples 
from the Galaxy, LMC or SMC.  

Earlier publications of data from our Local Group sample did 
not have the benefit of the periods determined from the SAAO, 
but we are now in a position to compare our Local Group 
carbon stars directly to the other samples.  
Figure~\ref{f.p_69} includes data for carbon stars in Fornax, 
Sculptor, and Leo~I, along with the comparison sample from 
the Galaxy, LMC, and SMC.  The overall dependency of dust 
content with pulsation period is unchanged.  The new Fornax 
data appear to follow the same dependency, although the 
period coverage is smaller.  This similarity is consistent 
with our revised metallicity.  The figure includes a line 
fitted to all of the Galactic, Magellanic, and Fornax data:
\begin{equation}
[6.4]-[9.3] = -0.227 + 0.00169 \, \, P \, {\rm (days)}.
\end{equation}

Interestingly, the five spectra from Sculptor and Leo~I all 
lie {\it below} the fitted line in Figure~\ref{f.p_69}.  The 
mean difference for these five is $-$0.185 magnitudes, with 
an uncertainty in the mean of 0.047 magnitudes.  The 
comparison data, including Fornax, show a standard deviation 
of 0.209 magnitudes about the fitted line, and with a sample 
of 121 objects with periods, the uncertainty in the mean is 
0.019 magnitudes.  Adding the uncertainties in quadrature 
gives an uncertainty in the difference between the fitted 
line and the data from Sculptor and Leo~I of 0.051 
magnitudes, making the difference between them and the other 
samples 3.6~$\sigma$.  

In contrast, the mean difference between the Fornax data and
the fitted line in Figure~\ref{f.p_69} is only 0.018 
magnitudes.  Comparing this difference to a standard 
deviation of 0.241 and an uncertainty in the mean of 0.098 
magnitudes shows that Fornax follows the Magellanic and 
Galactic samples.

The difference for Sculptor and Leo~I is statistically 
significant.  The carbon stars in these two galaxies almost 
certainly do not belong to the same population as Fornax, 
the SMC, the LMC, and the Galaxy ($p$ value = 0.00014).  
Nonetheless, we have only five deviant spectra, further
verification with larger samples would be helpful.
While we have revised previous estimates of the metallicity 
of the carbon stars in Fornax upward to Magellanic values and 
the metallicity in Sculptor from $-$1.4 to $-$1.0, the Leo~I 
sample restores the range of [Fe/H] sampled in this analysis
down to $\sim$ $-$1.35.  We conclude that for the most 
metal-poor stars in our sample, the impact of the initial 
metallicity of the star on its future dust production as a 
carbon-rich AGB star is large enough to be noticeable in the 
infrared.

\subsection{Silicon carbide dust emission\label{s.sic}} 

\begin{figure*} 
\includegraphics[width=7.0in]{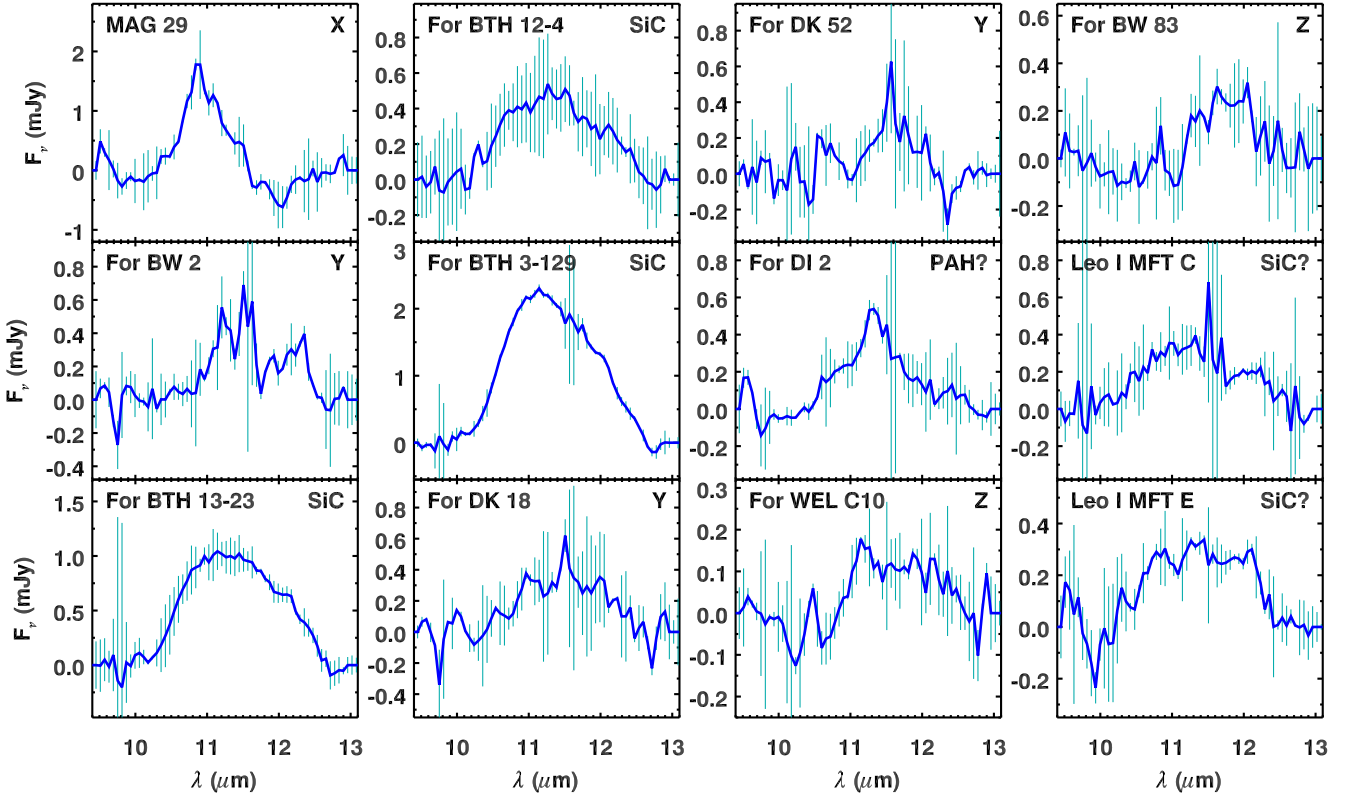}
\caption{The extracted spectral features in the vicinity of
the SiC dust emission feature for the 12 spectra where the
extracted flux has a S/N ratio $>$ 2.5, along with 
identifications of the features where possible.  ``X'' refers 
to the apparent feature in the spectrum of MAG~29 centered at 
10.9~\mum\ and most likely due to the continuum between 
molecular absorption bands.  ``Y'' labels features with a 
central wavelength at $\sim$11.5~\mum, and ``Z'' labels 
features centered to the red, 
$\sim$11.8--11.9~\mum.\label{f.sic}}
\end{figure*}

\subsubsection{In the dwarf spheroidals} 

The SiC feature at $\sim$11.3~\mum\ is generally weak in this
sample.  Only three sources show an unambiguous SiC feature,
although two more have spectral structure consistent with it.
In Table~\ref{t.dust}, seven features have a
S/N ratio $<$ 2.5.  For these, the central wavelength is 
omitted because it is meaningless; their continuum-subtracted 
spectra are essentially noise in this region.  
Figure~\ref{f.sic} illustrates the extracted features in the 
vicinity of 11.3--11.5~\mum\ for the remaining 12 sources 
which have an extracted strength with a S/N ratio $>$ 2.5.  

The five probable or possible SiC features all have central
wavelengths of 11.2--11.3~\mum.  Three of these are in 
Fornax (BTH 3-129, 12-4, and 13-23), and they were all
discussed previously by \cite{mat07}.  The other two are in
Leo~I.  In both cases, the emission profile is not a perfect
match for SiC, but given the noise, SiC is the most likely
explanation.

Fornax~DI~2 has an emission feature which peaks at 11.3~\mum\
and has a width and shape consistent with the out-of-plane
C--H solo bending mode in polycyclic aromatic hydrocarbons
(PAHs).  The problem with this assignment is that no other 
PAH features can be identified with any confidence.  There 
is a hint of the 8.6~\mum\ feature, but noise masks the 
6.2~\mum\ feature, and acetylene absorption would hide the 
7.5--7.9~\mum\ emission complex.  No 12.7~\mum\ feature is
apparent.

Other features in Figure~\ref{f.sic} can be placed in three
groups, labelled ``X'', ``Y'', and ``Z''.  

MAG~29 has the sole ``X'' feature, an apparent emission 
feature centered at 10.9~\mum, but its strength amounts to 
only 3\% of the continuum integrated over the same wavelength 
range.  Given the strong molecular absorption bands in this 
spectrum (\S~\ref{s.gas}), we suspect that the apparent peak 
at $\sim$11~\mum\ is simply the continuum between molecular 
absorption bands.  C$_3$ at 10~\mum\ might explain the drop 
to the blue side of this ``feature'' \citep{zij06}.  The drop 
to the red could be due to the broad wings of the C$_2$H$_2$ 
band centered at 13.7~\mum, which grow wider for higher gas
temperatures.

Three spectra show what we are calling the ``Y'' feature,
which peaks at $\sim$11.5~\mum\ and is sharper and more 
symmetric than the SiC feature.  These features are weak, and 
the spectra have poor S/N ratios, preventing any conclusive 
statements about their carrier.  Nonetheless, it should
be noted that graphite produces an emission feature in this
spectral range \citep{dl84,ld93}.  The absence of this 
feature in the spectrum of IRC +10216 \citep{mr87} and other 
Galactic carbon stars led to the currently favored model where 
amorphous carbon, and not graphite, dominates the dust around
carbon stars.  The 11.5~\mum\ graphite feature arises from
C--C displacements between graphene sheets, and in laboratory
data it is exceptionally narrow due to the regular spacing 
and large extent of these sheets.  Smaller sheets and 
irregularities in the lattice structure would fatten the
feature, but whether or not it would have a shape like the
observed ``Y'' feature is unknown.  The presence of graphite 
in carbon-rich dust shells is an interesting possibility, but
given the limited quality of the extracted features, any 
further speculation is unwarranted.  

Two spectra show possible emission features in this range
which peak further to the red, in the 11.8--11.9~\mum\ 
region.  We are calling this the ``Z'' feature.  As with the 
``Y'' feature, though, the data are noisy.  One could argue 
that the feature in the spectrum of For~WEL~C10 is a very 
noisy example of SiC emission, while the feature in 
For~BW~63 could be a noisy ``Y'' feature.

\subsubsection{Comparing the samples} 

\begin{figure} 
\includegraphics[width=3.25in]{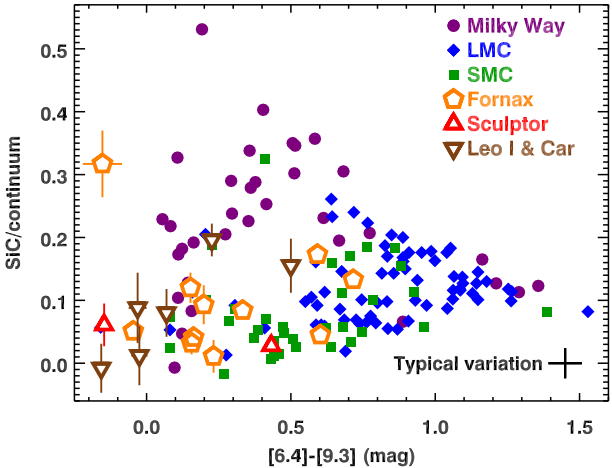}
\caption{The strength of the SiC dust emission at 
$\sim$11.3~\mum, relative to the underlying continuum from
star and amorphous carbon dust, plotted as a function of 
[6.4]$-$[9.3] color.  The error bars labelled ``Typical 
variation'' indicate the expected changes in spectral
properties over a pulsation cycle of the central star.  The
uncertainties are plotted for the program stars, but they 
are generally smaller than the plotting symbols.  The 
Galactic carbon stars follow a different sequence than most 
of the carbon stars in the other galaxies.
The one data point from Fornax with [6.4]$-$[9.3] = $-$0.15 
and SiC/cont.\ = 0.32 is For~DI~2, which appears to show 
emission from PAHs and not SiC at 
$\sim$11.3~\mum.\label{f.69sic}}
\end{figure}

Figure~\ref{f.69sic} plots the strength of the SiC dust 
emission, normalized to the underlying continuum as a 
function of the [6.4]$-$[9.3] color.  The SiC dust strength 
is integrated between 10.1 and 12.5~\mum\ and divided by the 
total ``continuum'' emission in the same interval, where 
``continuum'' is the combination of emission from amorphous 
carbon dust and the central star.  

\cite{slo06} found that the relative strength of the SiC 
emission feature decreased as the metallicity of the sample
decreased.  In Figure~\ref{f.69sic}, two different sequences 
of relative SiC strength vs.\ total dust content are 
apparent, with a clear bimodality at colors of 
$\sim$0.2--0.6.  In this range, Galactic stars dominate the 
upper sequence, with not one falling on the lower sequence.  
Once [6.4]$-$[9.3] exceeds $\sim$0.6, the sequences begin to 
merge, although the difference between the Galactic and 
Magellanic samples is still evident.


Generally, the carbon stars in the dwarf spheroidals follow 
the lower sequence in Figure~\ref{f.69sic}.  Fornax~DI~2 
(with [6.4]$-$[9.3] = $-$0.15 and SiC/cont.\ = 0.32) is the 
most significant exception, showing an emission feature at 
11.3~\mum\ with strong contrast to the continuum but 
virtually no other dust.  This emission feature looks more
like PAHs than SiC (see Figure~\ref{f.sic}), but as discussed
above, that identification is problematic and uncertain.  The 
blue [6.4]$-$[9.3] color indicates that For~DI~2 is virtually 
naked, making SiC dust unlikely.  Another major exception is 
Leo~I~MFT~E ([6.4]$-$[9.3] = 0.23, SiC/cont.\ = 0.20).  In 
this case, the feature is most likely SiC and the exception 
appears to be real.

Three carbon stars in dwarf spheroidals appear in 
Figure~\ref{f.69sic} with SiC/cont.\ $>$ 0.13 and 
[6.4]$-$[9.3] $>$ 0.45.  These objects are (left to right in
Figure~\ref{f.69sic}):  Leo~I~MFT~C, For~BTH~3-129, and
For~BTH~13-23.  While the apparent SiC emission in 
Leo~I~MFT~C is somewhat noisy, its profile is consistent with 
SiC.  The identifications of the features in the two Fornax 
spectra are firm, due to the strength and profiles of these 
features.  We see no obvious characteristics to distinguish 
these sources from the others with weaker SiC features, much
as \cite{slo06} found when investigating the five SMC sources
in the same region.  None of these eight stand out in terms 
of pulsation period, luminosity, or any other identified
property.  Why they have stronger SiC features than other
sources from the same galaxies remains unknown.

The sources in Fornax show SiC strengths similar to the 
Magellanic sources, fully consistent with the metallicities 
we believe they formed with.  The general lack of dust in the
three Carina sources places them in the lower left corner of 
Figure~\ref{f.69sic}, where the two tracks converge, giving
us little new insight.  The Leo~I sources, however, are a bit 
of a surprise, with one of the three on the upper sequence.  

\subsection{Acetylene gas absorption\label{s.gas}} 

\begin{figure} 
\includegraphics[width=3.25in]{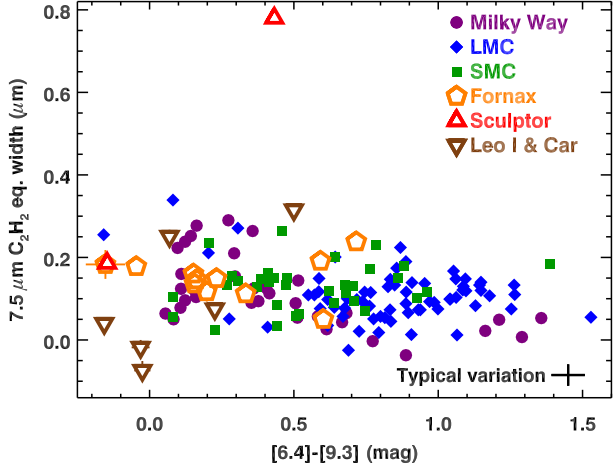}
\caption{The equivalent width of the absorption band from 
C$_2$H$_2$ at 7.5~\mum\ as a function of the [6.4]$-$[9.3] 
color.  The data point in the top center is MAG~29 in 
Sculptor.  This spectrum has a 7.5~\mum\ band more than twice 
as deep as any other measured.  The general trend of stronger
absorption bands with decreasing metallicity is most apparent
for 0.5 $<$ [6.4]$-$[9.3] $<$ 1.0.\label{f.69_75}}
\end{figure}

Table~\ref{t.gas} presents the equivalent widths of the
acetylene (C$_2$H$_2$) bands centered at 7.5~\mum\ and 
13.7~\mum.  As with Table~\ref{t.dust}, we do not quote the 
central wavelength if the S/N ratio of the equivalent width 
is less than 2.5.  In addition, we omit it if the 
uncertainty in the wavelength exceeds 0.2~\mum\ at 7.5~\mum\
or 0.10~\mum\ at 13.7~\mum.  In the cases with no
central wavelength in Table~\ref{t.gas}, the equivalent
width of the feature plus the uncertainty could be
considered as an upper limit.


The 7.5~\mum\ band arises from the P and R branches of the 
$\nu_4^1 + \nu_5^1$ transitions and often presents a 
double-troughed structure with a central peak at 7.5~\mum.  
The sharp absorption feature at 13.7~\mum\ arises from the 
Q branch of the $\nu_5$ transition, primarily the fundamental 
mode, but with some contribution from higher overtones.  The 
P and R branches extend this feature to cover the 
$\sim$13--15~\mum\ range, but our lack of LL coverage 
prevents us from measuring the full band.  Table~\ref{t.gas} 
presents the equivalent width of just the Q branch at 
13.7~\mum.  \cite{mat06} found that HCN, which often produces 
bands in the immediate vicinity of the C$_2$H$_2$ bands in 
Galactic carbon stars, was absent in the spectra of carbon 
stars in the LMC.  This conclusion also applies to carbon 
stars in the SMC \citep{slo06}.  No evidence of HCN appears 
in the Local Group spectra presented here, but we note that 
limited S/N ratios and wavelength coverage at 14~\mum\ 
prevent us from drawing any firm conclusions.

The wavelengths in Table~\ref{t.mm} were used for all
features except one, the 7.5~\mum\ band in MAG~29.  That band 
is so strong that the continuum wavelengths fall in the wings 
of the absorption, forcing us to shift them to 
6.14--6.44~\mum\ on the blue side and 9.12--9.52~\mum\ to the
red.  It is also possible that the extension of the
absorption to longer wavelengths is due to another molecule,
but that molecule would probably be another hydrocarbon and
thus trace the same molecular hydrocarbon mix as the 
acetylene feature.

We have detected a 7.5~\mum\ band in most of our spectra.
The exceptions include all three carbon stars in Carina as
well as For~DK~52.  All of the spectra with detections show 
a clear minimum in the wavelength range covered by the band, 
although the noise can be considerable in some cases, due 
most likely to mismatches in the individual nod spectra and 
non-gaussian contributions which become a problem at these 
low flux levels.

The narrow 13.7~\mum\ band is more difficult to detect.  In
addition to the lower signals at these wavelengths, only two
of the spectra have LL data, leaving only three data points
in SL for fitting the red continuum for the rest of the 
spectra, since we do not use SL past 14.17~\mum.  
Nonetheless, we still detect a 13.7~\mum\ band in seven of 
the 19 spectra, an impressive result given the sub-mJy 
strengths of most of the spectra at 14~\mum.

All of our sources in Sculptor, Fornax, and Carina have been
confirmed as carbon stars in the literature.  \cite{hel10}
describe two of the three targets in Leo~I, MFT~C and A, as 
probable carbon stars.  MFT~C shows a convincingly strong
7.5~\mum\ absorption band, and the 11.2~\mum\ emission 
feature, while noisy, shows a shape consistent with SiC,
allowing us to confirm its carbon-rich nature.   MFT~A also
has a strong 7.5~\mum\ band, although it is noisy.  Given
the lack of any obvious features at 11.3 and 13.7~\mum, 
this object remains an unconfirmed carbon star.
The chemistry of MFT~E is less clear in the literature.  The
IRS spectrum shows what is best described as a weak, but
noisy, 7.5~\mum\ acetylene band and an emission feature at
11.3~\mum\ consistent with SiC, which leads us to treat 
it as a probable, but unconfirmed carbon star.


We will focus on the 7.5~\mum\ absorption band from 
C$_2$H$_2$ due to its better S/N ratio.  
Figure~\ref{f.69_75} plots its equivalent width vs.\ 
[6.4]$-$[9.3] color, allowing us to compare the molecular 
band strength at similar overall dust contents.  The 
differences between the samples are most apparent in the 
color range 0.5--1.0, with a clear trend in increasing band 
strength from the Galaxy to the LMC to the SMC.  Again, the 
three equivalent widths from Fornax are generally consistent 
with the Magellanic sample.  Two of the three are among the 
strongest bands in this color range.  

The data from the other dwarf galaxies are outside this color 
range, but a couple of comments are in order.  First, MAG~29 
in Scl has an equivalent width more than twice as strong as 
anything else in the sample.  \cite{men11} noted that the IRS 
spectrum was obtained at maximum luminosity, which might 
account for the strong acetylene absorption.  Even if this 
band varied by a factor of two over a pulsation cycle, it
would still be stronger when at its minimum than in any other 
spectrum considered here.  Second, two of the three stars in 
Leo~I are among the strongest absorbers at 7.5~\mum.  The
strong acetylene absorption in MAG~29 and in Leo~I is
consistent with the metal-poor nature of Sculptor and Leo~I.


\subsection{Metallicity diagnostics} 

\begin{deluxetable}{lrcc} 
\tablenum{12}
\tablecolumns{3}
\tablewidth{0pt}
\tablecaption{Mean SiC and acetylene strengths}
\label{t.mean}
\tablehead{
  \colhead{Galaxy} & \colhead{Number} & 
  \colhead{$<$SiC/cont.$>$\tablenotemark{a}} &
  \colhead{$<$$7.5~\mum\ $EW$>$\tablenotemark{a}}
}
\startdata
Milky Way &  8 & 0.29 $\pm$ 0.07 & 0.06 $\pm$ 0.04 \\
LMC       & 23 & 0.11 $\pm$ 0.07 & 0.08 $\pm$ 0.05 \\
SMC       & 12 & 0.08 $\pm$ 0.06 & 0.12 $\pm$ 0.05 \\
For dSph  &  3 & 0.12 $\pm$ 0.06 & 0.16 $\pm$ 0.10 \\
\enddata
\tablenotetext{a}{In the range 0.5 $<$ [6.4]$-$[9.3] $\le$ 0.8.}
\end{deluxetable}

\begin{figure} 
\includegraphics[width=3.25in]{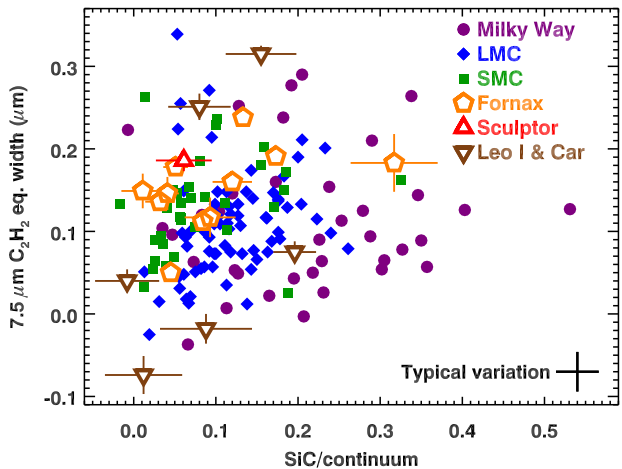}
\caption{The equivalent width of the acetylene band at 
7.5~\mum\ plotted as a function of the SiC/continuum emission 
ratio.  MAG~29 is missing from this plot because it is so far
above the rest of the data, with EW = 0.78~\mum\ and 
SiC/cont.\ = 0.03.  \cite{lag08} proposed this comparison as 
a metallicity diagnostic (their Fig.~11).\label{f.75_sic}}
\end{figure}

Table~\ref{t.mean} summarizes the mean and standard deviation
of the strength of the SiC emission (normalized to the 
continuum) and the equivalent width of the 7.5~\mum\ 
acetylene absorption band.  Both of these quantities vary 
with [6.4]$-$[9.3] color, and the various samples considered 
are incomplete at some colors.  Consequently, we limit the 
comparison to the range 0.5 $<$ [6.4]$-$[9.3] $\le$ 0.8.
From the Milky Way to the LMC to the SMC, the trend of
decreasing SiC strength and increasing C$_2$H$_2$ strength
is clear.  It is possible that the overlapping spreads in
the data arise from a range of metallicities within the
three galaxies.  As described above (\S~\ref{s.metal} and
\ref{s.mm}), we suspect that the metallicity of Fornax is 
closer to Magellanic than to the other dwarf spheroidal 
galaxies considered here.  We have only three spectra in
this color range, and they give ambiguous results.  The SiC
feature suggests a similarity to the LMC, but the spread in
acetylene strengths limits our conclusions.

\cite{lag08} proposed plotting the two quantities in 
Table~\ref{t.mean} against each other to diagnose 
metallicity.  Figure~\ref{f.75_sic} follows that lead.
While the sample from each galaxy shows considerable scatter, 
the gradient is unmistakable, with the Galaxy dominating in
the lower right, the LMC in the middle, and the SMC in the
upper left.  The Fornax spectra are distributed much like
the SMC spectra.  As explained above, the one discrepant 
point in Fornax is For~DI~2, which might have PAHs, not SiC, 
in its spectrum.  Overall, Figure~\ref{f.75_sic} reinforces 
our suspicions about Fornax and suggests that it and the SMC 
have similar metallicity distributions.  

The Sculptor data keep to the metal-poor side of the diagram,
with MAG~29 literally off the charts (not plotted, with 
SiC/cont.\ = 0.03 and EW = 0.78~\mum).  This behavior is 
fully consistent with the narrow and metal-poor MDF for 
Sculptor.  The three Carina spectra are clustered in the 
lower left, where metallicity is indeterminant.  The Leo~I 
data are more of a puzzle.  Two of the three spectra are in 
the upper left, where we would expect them from the 
metallicity of their host galaxy, but Leo~I~MFT~E lands 
squarely among the metal-rich spectra.  This might indeed 
indicate a higher initial metallicity, but it may also 
suggest that a bit of caution is warranted with these data.  
One alternative explanation is that this spectrum is affected 
by strong C$_3$ absorption at 10~\mum, which would push our 
continuum fit downward and enhance the apparent strength of 
the SiC emission.

\section{Discussion} 

\subsection{Carbon dust content and metallicity~\label{s.cz}} 

The most surprising result of this study is that the carbon 
stars in Sculptor and Leo~I seem to have less dust around 
them than their counterparts in more metal-rich galaxies.
Previous studies of carbon stars in the Galaxy, the LMC, and 
the SMC showed that the amount of circumstellar dust, as
measured by the [6.4]$-$[9.3] color, followed the same 
relationship with pulsation period of the star, despite their 
differences in metallicity.  The six carbon stars in Fornax 
with known pulsation periods also follow this relation.  As
explained in \S~\ref{s.mm}, we believe that these stars have 
metallicities similar to those in the LMC and SMC.  The two 
carbon stars in Sculptor, for which we estimate 
[Fe/H]$\sim$$-$1.0, and the three carbon stars in Leo~I, with 
[Fe/H]$\sim$$-$1.35, are shifted downward on average about 
0.19 magnitudes in Figure~\ref{f.p_69}.  This difference 
corresponds to a factor of two in dust production 
($\dot{D}$, Equation~1).

The scatter in Figure~\ref{f.p_69} is considerable.  The 
standard deviation about the fitted line corresponds to a 
range in $\dot{D}$ of a factor of 2.2 (up or down).  The 
cause of this wide envelope around the general proportionality 
of dust content to pulsation period is an open question.  A 
more careful look at the Magellanic data rules out luminosity 
(Sloan et al., in preparation).  Metallicity also does not 
work, even though we expect a fairly broad range of 
metallicities within each of the Milky Way, LMC, and SMC.  If 
metallicity were responsible, it would lead to a measurable 
shift in the mean dust content between each galaxy, which we
do not see.

The data leave us to conclude that for [Fe/H] $\ga$ $-$0.7,
the amount of carbon-rich circumstellar dust shows no 
relation to the metallicity.  The remainder of the
discussion focuses on the implications.  As stated in 
\S~\ref{s.dust} above, the carbon dust content is measured 
by the [6.4]$-$[9.3] color, with no assumptions about outflow 
velocity or gas-to-dust ratio.  

\subsection{C/O ratio and metallicity~\label{s.co}} 

For a carbon star, the dust content depends primarily on the 
abundance of C, which it produces via the 3-$\alpha$ sequence 
and dredges to its surface.  The free carbon available for 
dust production depends on what remains after CO consumes the 
available oxygen, so that  
\begin{equation}
\frac{C_{\rm free}}{O} = \frac{C}{O} - 1.
\end{equation}
In this equation and the ones that follow, ``C'' and ``O'' are
the fractional abundances of carbon and oxygen, by number.
Not all of the free carbon will form dust; some of it appears
in our spectra as acetylene, and other carbon-bearing 
molecules are possible.  Because our carbon stars are 
embedded in dust and their distances make them faint, they 
are difficult targets for the high-resolution optical 
spectroscopy usually used to determine C/O ratios.  While
waiting for that problem to be solved, we can still
estimate what C/O ratios we should expect.  We can write
\begin{equation}
\frac{C}{O} = \frac{ C_i + \delta C } {O} =
  \left( \frac{C}{O} \right)_i + 
  \left( \frac{\delta C}{O_i} \right),
\end{equation}
where $C_i$ is the initial carbon abundance, $\delta C$ is
the change in abundance due to dredge-up, and we have
assumed that the oxygen abundance does not change from its 
initial level.

Observations of the C/O ratio in dwarf galaxies reveal a
decrease in C/O with decreasing metallicity \citep[][where 
O/H serves as a proxy for metallicity]{gar95,kob98}.  This is 
the initial C/O ratio prior to any dredge-up on the AGB.  
Data from stellar spectroscopy in the Galaxy compiled by 
\cite{chi99,chi03} show no significant trend in [C$_i$/Fe] 
with [Fe/H].  The scatter is substantial, $\sim$0.2 dex, and 
the point is not without controversy, but we will assume no
systematic relation between [C$_i$/Fe] and [Fe/H].  Similar 
spectroscopy by \citet[][and references therein]{mel02} shows 
that a steadily increasing [O/Fe] ratio in more metal-poor 
samples is the culprit behind the changing (C/O)$_i$ ratio.  
We estimate that in the range $-$1.5 $\la$ [Fe/H] $\le$ 0.0, 
[O/Fe] increases 0.25 dex for every decrease of 1 dex in 
[Fe/H].  The scatter about this trend is $\sim$0.1 dex on 
average.  Applying this trend to the above equation gives:
\begin{equation}
\frac{C}{O} = \left( \frac{C}{O} \right)_{\odot} \left(
  10^{0.25 \, {\rm [Fe/H]} } + 
  \delta C \, 10^{-0.75 \, {\rm [Fe/H] }} \right).
\end{equation}
The first term in the right-most parentheses is just the 
initial C/O ratio, which decreases by a factor of 1.78 for 
every dex in [Fe/H].  

Let us assume that the amount of dredged up carbon ($\delta 
C$) does not vary significantly with metallicity.  The
arguments which follow will pursue the consequences and
show that in fact this assumption is probably invalid.

For the sake of argument, we will also assume that a typical
C/O ratio for a carbon star at solar metallicity is $\sim$1.1 
\citep[following][but see also Ohnaka et al.\ 2000]{lz08}.  
At solar metallicity, the initial C/O ratio of a star would 
be 0.54 \citep{asp05}.  To raise the C/O ratio to 1.1, 
$\delta C$ would have to be 0.56~O$_{\odot}$.  Keeping this 
quantity fixed with metallicity leads to a C/O ratio at 
analogous stages of AGB evolution of 1.4 at [Fe/H] = $-$0.3 
(i.e.\ in the LMC), 2.2 at [Fe/H] = $-$0.7 (SMC), and 3.5 at 
[Fe/H] = $-$1.0 (Sculptor).  

Dividing Equation (2) by (C/O)$_{\odot}$ leads to a relation
for the free carbon as a fraction of the total initial carbon 
at solar metallicity:
\begin{equation}
\frac{C_{\rm free}}{C_{\odot}} = 
   \left( \frac{C}{O} - 1 \right) 1.85 \times 10^{0.75 \, 
   {\rm [Fe/H]}}.
\end{equation}

For a star of solar metallicity, $C_{\rm free}$ = 0.19 
$C_{\odot}$ (in this particular case, $C_{\odot}$ is the 
amount of carbon the star formed with).  The corresponding 
values for the LMC, SMC, and Sculptor are 0.44, 0.68, and 
0.81.  It is instructive to consider the example of the SMC.  
Despite the fact that stars in the SMC form with only 20\% of 
the carbon in their Galactic counterparts, if they dredge up 
the same amount of carbon, they will have 3.6 times {\it 
more} carbon free to form dust.  This is a direct result of 
the reduced oxygen abundance, which limits how much carbon is 
sequestered as CO.  All things being equal, {\it we would 
expect to see more carbon-rich dust at lower metallicities}, 
and yet, our measurements of the carbon dust content remain 
relatively constant with metallicity.  We even see a possible 
dip at the metal-poor end of our sample.


\subsection{Acetylene, the carbon budget, and mass-loss triggers} 

Before giving up on our assumption that $\delta C$ does not
depend on metallicity, we should consider what might happen
to this excess of free carbon at low metallicity.  If the
dust content stays flat or even falls, some other reservoir 
must be absorbing the carbon.  SiC represents only trace 
amounts, but the strengthening acetylene bands at lower 
metallicity could be a solution.

Table~\ref{t.mean} shows that the mean strength of the
7.5~\mum\ acetylene band is twice as strong in the SMC than
in the Milky Way, although the spread of the data about 
these means is considerable.  If the 7.5~\mum\ acetylene band 
accurately traces the total carbon mass in gaseous 
hydrocarbon molecules, then it follows from 
Table~\ref{t.mean} that hydrocarbon gases sequester two times 
more carbon in the typical carbon star in the SMC compared to 
the Milky Way.  

While this is a significant amount of carbon in gaseous form,
it is not enough to account for the factor of 3.6 more free 
carbon expected in the SMC.  We know of no other likely 
reservoir for the carbon, leading us to conclude that the 
amount of dredged up carbon ($\delta C$) must in fact 
decrease at lower metallicity.  Neither the rate at which the 
triple-$\alpha$ sequence produces carbon nor the dredge-up 
efficiency should decrease at lower metallicity.  To reduce 
$\delta C$, it is necessary to terminate the dredge-up 
process progressively earlier on the AGB in more metal-poor 
stars.

Models by \cite{woi06} show that the opacity of carbon-rich 
dust is sufficient to drive the mass-loss process.  If a 
dredge-up event during a thermal pulse drives the C/O ratio 
high enough, then the jump in free carbon will lead to a 
pulse of dust formation which could quickly strip the 
envelope and bring the AGB evolution to a rapid end.  
\cite{lz08} proposed just such a scenario, although their
definition of the free carbon differs from ours.  The key
point is that the trigger is not the C/O ratio, but the
quantity of free carbon, modified, if seeds are important 
to the condensation process, by the abundances of elements
such as Ti and perhaps Si.  A spectroscopic census of the 
Sgr dSph galaxy (McDonald et al., in preparation) finds that 
the typical lifetime of a star as a carbon star may be similar 
to the time between thermal pulses.  Once a star develops a 
carbon-rich envelope, the formation of dust and the loss of 
mass accelerates quickly.

\subsection{Si, seeding, and abundance} 

Van Loon et al.\ (2008) explained the increasing acetylene 
band strengths in more metal-poor stars as a consequence of 
the lower abundances of elements like Ti.  The lower 
abundances would result in fewer available seeds like TiC 
for the condensation of amorphous carbon dust, making the 
process less efficient.

SiC could play a similar role as a condensation seed.  
\cite{slo06} suggested that the decreasing SiC emission at 
lower metallicity would arise naturally from decreasing Si
abundances.  \cite{lag07} and \cite{lei08} developed this
idea, explaining the different tracks of SiC emission 
strength with [6.4]$-$[9.3] color in Figure~\ref{f.69sic} 
in the different samples as an increasingly delayed 
condensation of SiC with decreasing metallicity.  Other 
seeds requiring heavy elements should behave similarly with 
decreasing metallicity.  Without these seeds at lower
metallicities, the less efficient formation of amorphous
carbon is likely to result in a greater fraction of the free
carbon tied up in simpler molecules like acetylene.

Observations of Magellanic planetary nebulae (PNe) by 
\cite{ber09} create one difficulty for the idea that the 
reduced SiC strength in metal-poor carbon stars arises from 
reduced Si abundances.  They found that strong emission 
features centered at $\sim$11~\mum\ in several spectra, and
they attributed these features to SiC.  We suspect that the
11~\mum\ feature in their PN spectra does not arise from
SiC, but instead from a different carrier.  The PN features
are triangular in shape and peak at $\sim$11.1~\mum, compared
to the more rounded SiC features which peak at 
$\sim$11.3~\mum.  Additionally, the PN features are usually 
associated with features at $\sim$16~\mum.  Both the 11 and 
16~\mum\ features are often seen in spectra showing the still 
unidentified 21~\mum\ feature \citep[][Sloan et al.\ in
preparation]{kra02}.

\subsection{Gas-to-dust ratio and outflow velocity\label{s.vout}} 

For evolved oxygen-rich stars, the gas-to-dust ratio should
depend strongly on metallicity, because the abundances of the 
elements which form oxygen-rich dust scale with metallicity 
\citep[e.g.,][]{vl00}.  The outflow velocity depends on the
radiation pressure on the dust and its coupling to the gas
due to collisions, and it should also vary with metallicity.

The situation for carbon stars is different, because they 
produce the carbon which drives the mass-loss themselves.  
The spectra in this paper show that if the dust content 
decreases with lower metallicity, the dependence is weak and 
noticeable only for the most metal-poor stars observed. 
Following the carbon budget leads us to the same conclusion 
as \cite{lz08}, that the free carbon serves as a trigger for 
the superwind phase.  

The lack of a strong dependence of carbon-rich dust content 
with metallicity makes it difficult to see how the gas-to-dust 
ratio or the outflow velocity should depend on metallicity.
If radiation pressure accelerates the dust, and collisions 
between dust grains and gas molecules accelerate the gas,
then the amount of dust would determine both outflow velocity
and total mass-loss rate, and as we have shown, the amount
of dust does not vary strongly with metallicity.
If pulsations drive the mass-loss process, then the mass-loss
rate will depend primarily on the pulsation period and
amplitude, which are relatively insensitive to metallicity.
Again, neither the outflow velocity nor the gas-to-dust ratio
should depend strongly on metallicity.

A rigorous test of these predictions awaits observations from
the Atacama Large Millimeter Array (ALMA).  Less sensitive 
telescopes can be used to study carbon stars in the Galactic 
Halo.  \cite{lag10} observed a sample of six carbon stars in 
the Halo and found that they have smaller outflow velocities 
than carbon stars in the Galactic disk.  \cite{lag12} obtained 
infrared spectra of four of these carbon stars with the IRS,
and they found evidence suggesting that halo Carbon stars 
also have high gas-to-dust ratios.  Thus, both outflow 
velocity and gas-to-dust ratio may depend on metallicity for 
carbon stars, after all, but some caution is in order.  
First, the metallicity of the observed carbon stars is 
uncertain; they may be ejected from the Galactic disk.  
Second, if the outflow velocity of the dust is higher than 
the gas, then the inferred gas-to-dust ratios may not differ 
from the disk sample as much.  The Halo carbon stars raise 
some interesting possibilities, but these initial results 
require follow-up.

The review of CO line observations in Galactic carbon stars 
by \cite{sch07} shows a range of possible outflow velocities,
from a few km/s to over 20 km/s, with higher outflow 
velocities generally associated with higher total mass-loss 
rates.  The relation between outflow velocity and total 
mass-loss rate is consistent with increasing amounts of 
circumstellar dust, which leads to more efficient 
acceleration of the circumstellar envelope.  Metallicity is 
not needed to explain the effect, even if the range of 
possible outflow velocities does inject a degree of caution 
when using Equation~(1) or interpreting the mass-loss rates 
presented in Table~\ref{t.dust}.

\section{Summary} 

The sample of carbon stars beyond the Magellanic Clouds 
observed by the IRS on {\it Spitzer} consists of 19 stars in
the Sculptor, Carina, Fornax, and Leo~I dwarf spheroidal 
galaxies.  The invidual targets were chosen based on their
red NIR colors or their pulsation properties.  The bolometric 
magnitudes of the two targets in Sculptor, their pulsation 
periods, and reference to recent modelling work lead us to 
revise their estimated metallicity ([Fe/H]) up to 
$\sim$$-$1.0.  Similarly, we suggest that the carbon stars
in Fornax have metallicities more similar to the LMC and
SMC than to the other dwarfs studied.  The mean metallicity 
in Leo~I and Carina are $\sim$ $-$1.4 and $-$1.7, 
respectively.

All carbon stars follow a general relation of increasing
dust content as the pulsation period increases.  Metallicity
does not show any influence down to [Fe/H]$\sim$$-$0.7, so
that despite their differences in initial metallicity, carbon
stars in the Galaxy, LMC, SMC, and Fornax with similar
pulsation periods produce similar quantities of dust.  The 
five stars observed in Sculptor and Leo~I produce less dust,
at a significance level of 3.6~$\sigma$.  The carbon stars in
Carina show no dust, but lack of pulsation periods prevent 
us from comparing them to the stars in the other galaxies.

The new carbon stars extend the previously detected trends
of decreasing emission from SiC dust and increasing 
absorption from acetylene gas as the metallicity falls to
even lower metallicities.  The carbon stars in Fornax show
SiC features and acetylene bands consistent with our
revised metallicity.

If the quantity of freshly produced carbon from the stellar 
interior to the photosphere does not depend on metallicity,
the lower initial abundances of oxygen in more metal-poor
stars would lead to substantially greater quantities of 
free carbon.  Yet, as our observations have shown, they do
not produce substantially larger amounts of amorphous carbon
dust.  While more metal-poor stars do produce more acetylene 
gas, it is not enough to account for the expected increases
of free carbon.  We conclude that some process must truncate
the dredge-up earlier in more metal-poor samples, quite 
likely a superwind triggered by a free-carbon threshold which
strips the star of its envelope and ends its evolution on the 
AGB.

Because the amount of circumstellar carbon-rich dust observed 
does not depend strongly on the metallicity, we should not
expect any strong dependencies of outflow velocity or 
gas-to-dust ratio on metallicity as well.

\acknowledgments

G.~C.~S. was supported by NASA through Contract
Number 1257184 issued by the Jet Propulsion Laboratory,
California Institute of Technology under NASA contract 1407.
Special acknowledgement is due to M.~Feast, P.~Whitelock, 
J.~Menzies, and all of their collaborators at the SAAO for 
their NIR study of the stars in our Local Group sample.  
These time-consuming observations have added immeasurably
to our {\it Spitzer} data.
This research has made use of NASA's Astrophysics Data
System, the Infrared Science Archive at the Infrared 
Processing and Analysis Center, which is operated by JPL,
and the SIMBAD and VIZIER databases, operated at the Centre 
de Donn\'{e}es astronomiques de Strasbourg.  (It is growing
difficult to picture how we did astronomy before these
wonderful services!)

\end{document}